\documentclass[reprint,nofootinbib,aps,pra,amsmath,amssymb]{revtex4-2}
\usepackage[T1]{fontenc}
\usepackage[colorlinks,citecolor=blue,linkcolor=blue,urlcolor=blue]{hyperref}
\usepackage{graphicx}
\usepackage{dcolumn}
\usepackage{bm}
\usepackage{bbm}
\usepackage{hyperref}
\usepackage{newtxtext}
\usepackage{newtxmath}
\usepackage{physics}
\usepackage{epsfig}
\usepackage{xcolor}
\usepackage[caption=false,font=footnotesize,justification=centering]{subfig}
\newcommand*{\Scale}[2][4]{\scalebox{#1}{$#2$}}%
\definecolor{green}{RGB}{20, 140, 40}

\begin{document}

\title{Quantum Advantage in Beam-Displacement Estimation \\ for Inter-Satellite Quantum-Secure Time Transfer}

\author{Ziqing Wang}%
 \email{ziqing.wang1@unsw.edu.au}
\affiliation{School of Electrical Engineering and Telecommunications, University of New South Wales, Sydney, NSW 2052, Australia}%
\author{Robert Malaney}%
 \email{r.malaney@unsw.edu.au}
\affiliation{School of Electrical Engineering and Telecommunications, University of New South Wales, Sydney, NSW 2052, Australia}%

\date{\today}%

\begin{abstract}
In this work, we investigate the practical feasibility of achieving a quantum advantage, in the context of inter-satellite quantum communication, of satellite position estimation. The investigation focuses on a quantum-enhanced scheme in which squeezing is added to the derivative mode of a Gaussian probe beam as a means of estimating photon-beam displacement. The performance of the scheme is compared against the standard quantum limit achieved by an optimal classical beam-displacement estimation scheme under a fair same-total-energy constraint.
We find that although the benefit provided by squeezing rapidly diminishes as the inter-satellite distance increases due to diffraction-induced beam spreading and subsequent truncation by the finite receiver aperture, a quantum advantage can still be expected over inter-satellite distances up to approximately 100 km under practical satellite configurations. This distance scale between satellites is anticipated for Starlink-type satellite constellations.
\end{abstract}

\maketitle

\section{Introduction}

Advanced communication networks in space operating under inter-satellite  classical optical communication are now a reality, most notably in the `Starlink' constellation~\cite{rm1,rm2}. Currently operating with approximately 10,000 Low-Earth Orbit (LEO) satellites, this constellation is predicted to grow to approximately 40,000 satellites in the near future~\cite{rm3}. Space-based quantum communication is also a reality, with satellite-ground quantum information transfer well established by the Micius spacecraft~\cite{MiciusRMP2022}. With these advances established, we open up the possibility of a quantum enabled version of Starlink in which the optical communication between the satellites carries not only classical information, but also quantum information. Among the numerous benefits of such a quantum space-based network, a specific application  would be a highly secure  time-transfer protocol known as Quantum-Secure Time Transfer
 (QSTT); a protocol that integrates Quantum Key Distribution  (QKD) with time transfer~\cite{MiciusTimeTransfer2020}. When coupled with enhanced timing attributes from entangled photon pairs, and optimal use of  \textit{a priori} shared secret keys for additional obfuscation, a high performance and highly secured QSTT system can be developed~\cite{rm6}.

A fundamental assumption of QSTT is that the round-trip time must be measurable to one of the two terminals within a prescribed alert limit~\cite{SecureSyncReq2018, MiciusTimeTransfer2020}. This means that the distance between the terminals must be obtained securely in advance, which, in turn, requires their relative positions in real-time to be precisely known.
In the context of inter-satellite optical communication, relative navigation and tracking techniques are commonly used to determine the current relative position vector between two terminals and to predict their future positions from measurements of relative range and line-of-sight direction~\cite{MiciusTimeTransfer2020}.
It is also possible that quantum position verification techniques can be applied in this context (see~\cite{rm8} for the history of this application and~\cite{rm9a,rm9} for recent experimental advances in the area).
In optical tracking scenarios, a laser probe beam may be transmitted between the terminals, enabling the receiving terminal to infer tracking information from the received optical field. By measuring the transverse displacement of the received beam relative to the known geometric center of the receiving aperture (or the local wavefront tilt with respect to the receiver optics), real-time corrections to the estimated relative positions can be derived and applied. 
The precision of such a measurement directly affects the reliability and security of the high-precision time-information distribution in the satellite network. 

Conventionally, beam displacement (as well as its conjugate variable, beam tilt) is inferred using direct-detection techniques, such as those based on photodetector arrays. Although straightforward to implement and widely used for optical tracking, these techniques generally do not realize the optimal quantum-noise-limited measurement required to reach the standard quantum limit (SQL) for beam-displacement estimation. By contrast, homodyne detection with a TEM$_{10}$-mode local oscillator has been shown to provide an optimal measurement of small displacement and tilt of a Gaussian beam without non-classical resources. Specifically, TEM$_{10}$ homodyne detection achieves a higher efficiency than direct-detection techniques and can saturate the SQL (see, e.g.,~\cite{SqueezingHomodyneTheory2004,TEM10_HomodyneOptimal2006,DetectionNoiseModesOptimal2006}).

Multimode Gaussian quantum light has been widely regarded as a powerful resource for practical, ultra-high-precision parameter estimation because such light can be generated at very high photon numbers, with substantial squeezing and entanglement shared across many modes (see discussions in, e.g.,~\cite{UltimateSensitivity2012}). 
For the quantum-enhanced measurement of beam displacement and tilt, multimode Gaussian squeezed light allows both direct-detection and homodyne-detection schemes to achieve a level of precision beyond the SQL~\cite{SqueezingHomodyneTheory2004,TEM10_HomodyneOptimal2006}. 
With such squeezed light, extensive theoretical (e.g.,~\cite{SmallDisplacement2000,UltimateQL_General2003}) and experimental (e.g.,~\cite{SplitDetectExp2002,QuantumLaserPointer2003,SqueezingQuadrant2004}) efforts have demonstrated beyond-SQL precision in the measurement of the displacement and tilt of a Gaussian beam using direct detection. 
Beyond-SQL measurements of small displacement and tilt of a Gaussian beam have also been demonstrated experimentally using multimode Gaussian squeezed light and homodyne detection (e.g.,~\cite{DetectionNoiseModesOptimal2006,TEM10_HomodyneOptimal2006}). 
It has been subsequently shown that the most economical way to reach the ultimate limit for a given resource of multimode Gaussian quantum light is to place the most squeezed state available in a well-defined light mode~\cite{UltimateSensitivity2012}.

Despite these advances, it remains unclear whether multimode Gaussian quantum light can retain quantum advantage in beam-displacement estimation for inter-satellite QSTT, where the receiving satellite has a finite, hard-edged aperture. 
Such an aperture truncates the probe beam after propagation, causing diffraction loss and altering the parameter-sensitive spatial modes at the receiver. 
The foundational studies~\cite{SqueezingHomodyneTheory2004,TEM10_HomodyneOptimal2006,DetectionNoiseModesOptimal2006,UltimateSensitivity2012,SmallDisplacement2000,UltimateQL_General2003,SplitDetectExp2002,QuantumLaserPointer2003,SqueezingQuadrant2004} largely considered idealized settings with negligible aperture-induced spatial truncation. 
More recent work has analyzed diffraction loss and parameter-sensitive receiver modes using a Gaussian soft-aperture model (e.g.,~\cite{SoftAperture2026}), but the corresponding receiver-mode structure and attainable precision for a hard-edged aperture in inter-satellite QSTT remain to be established.
In particular, it remains unclear how much quantum advantage is attainable in this setting relative to an SQL-attaining classical beam-displacement estimation scheme under the same transmitted-energy and aperture constraints. These questions are particularly relevant to inter-satellite QSTT, where moving space-based platforms require precise relative positioning and tracking under stringent size, weight, and power constraints.

To the best of our knowledge, this work provides the first investigation of the practical achievability of quantum advantage from the use of multimode Gaussian squeezed light in beam-displacement estimation for inter-satellite QSTT (and other inter-satellite communication applications). Specifically, we examine whether this advantage persists when a finite, hard-edged receiver aperture truncates the probe beam after propagation.
Here, the quantum advantage is quantified as the ratio between the minimum displacement-estimation uncertainty of an optimal classical scheme (which is known to achieve the SQL) and that of the quantum-enhanced scheme, as evaluated using the Quantum Fisher Information (QFI) formalism under the constraint of equal total transmitted energy. The optimal classical scheme transmits a coherent Gaussian probe beam, whereas the quantum-enhanced scheme adds squeezing to the TEM$_{10}$ mode to form a multimode probe beam. Our results show that although the benefit provided by squeezing rapidly diminishes as the inter-satellite distance increases, a quantum advantage can still be expected over inter-satellite distances of up to 100 km.

\section{System Model}\label{Sec:SysModel}

\begin{figure*}[hbt!]
    \centering
    \includegraphics[width=0.9\linewidth]{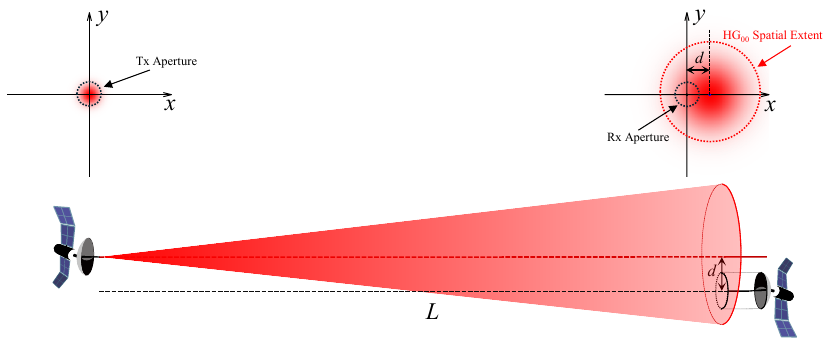}
    \caption{System model for optical tracking via beam-displacement estimation for inter-satellite QSTT. Note that the beam displacement $d$ is exaggerated for illustration purposes.}
    \label{fig:SysModel}
\end{figure*}
\twocolumngrid
We consider a typical system model (illustrated in Fig.~\ref{fig:SysModel}) for optical tracking  where one satellite estimates the transverse displacement $d$ of the laser probe beam transmitted from another satellite in order to acquire the real-time correction to the relative line-of-sight direction of that satellite.
For exposition, we focus on a typical inter-satellite configuration, where the transmitter and receiver aperture radii are both set to $a=15\,\text{cm}$, and the probe beam has a coherent component in the fundamental Gaussian ($\text{HG}_{00}$) mode with beam-waist radius $\omega_0=5\,\text{cm}$. Such a configuration resembles that employed in the Micius quantum experiment~\cite{SatQKD} and allows for negligible aperture truncation for $\{\text{HG}_{00},\text{HG}_{10},\text{HG}_{01}\}$ modes at the transmitter. We adopt the typical optical wavelength $\lambda = 1064\,\text{nm}$. The inter-satellite distance is denoted by $L$. We discuss later the effect of varying these settings.

\subsection{Assumptions}\label{Sec:Assumptions}
We adopt the following spatial profile for the $\text{HG}_{00}$ mode of the probe laser beam:
\begin{equation}\label{Eq:GaussianBeam}
    u_0(x,y,z)=\frac{1}{\sqrt{\pi w^2(z)}} e^{-\frac{x^2+y^2} {2 w^2(z)}}, 
\end{equation}
where $(x,y)$ denote the transverse coordinates, $z$ denotes the longitudinal coordinate, and $w:=w(z)=\frac{\omega_0}{\sqrt{2}} \sqrt{1+\left(\frac{z}{z_{\text{R}}}\right)^2}$ denotes a scaled beam radius\footnote{In this work, we use $w$ (without subscript) to denote the scaled beam radius, and we use $\omega_0$ to denote the beam-waist radius. In other words, $\omega_0=\sqrt{2}w(0)$.} 
with $z_{\mathrm{R}}=\pi \omega_0^2 / \lambda$ being the Rayleigh distance. 
For simplicity, we ignore any phase term (e.g., the Gouy phase) in all spatial modes. We also note that the definition in Eq.~(\ref{Eq:GaussianBeam}) is slightly different from the common convention in the field of optics due to the presence of the factor 2 in the denominator of the exponential function.

Without loss of generality, we assume that the laser probe beam is displaced along the $x$ axis (with displacement $d$)\footnote{We consider transverse beam displacement only and assume the beam tilt at the receiver to be zero. For an analysis of beam tilt, the reader is referred to~\cite{TEM10_HomodyneOptimal2006} where it is shown how to measure beam displacement at the SQL even in the presence of tilt. A similar analysis to that provided here could also be applied to the consideration of beam tilt.}
\begin{equation}\label{Eq:DisplacedGaussianBeam}
\begin{aligned}
    u_0[d]:=&u_0[d](x,y,z)\\
    =&u_0(x-d,y,z)=\frac{1}{\sqrt{\pi w^2(z)}} e^{-\frac{(x-d)^2+y^2} {2 w^2(z)}},
\end{aligned}
\end{equation}
is the displaced Gaussian beam before the truncation of the circular receiving aperture $A:=A(x,y)$ (which will be explicitly defined by Eq.~(\ref{Eq:Ap}) in Sec.~\ref{Sec:Aperture}). For clarity,  we will omit the spatial coordinates $(x,y,z)$ whenever possible. 

For practical reasons,  we assume displacement estimation is integrated into an inter-satellite system with precise point-ahead capability. This allows us to employ the common local-estimation assumption that the displacement to be estimated is very small, i.e., $d\rightarrow 0$, $d \ll w(z)$, and $d \ll a$. Under this approximation, the mean field, represented by the displacement vector in phase space, is assumed to not depend explicitly on the beam displacement.

\subsection{Displacement Estimation Schemes}
We introduce the two practical displacement estimation schemes considered and then define the quantum advantage. 
In addition, we adopt quantum parameter estimation theory and the QFI framework as detailed in~\cite{GaussianQFI2024}  to investigate the performance of the considered schemes. For completeness, this framework is summarized in Appendix~\ref{Sec:QET}.

\subsubsection{Optimal Classical Scheme}\label{Sec:OptCS}
We adopt an optimal classical scheme for displacement estimation as a benchmark. In this scheme, the laser probe beam is a single-mode field prepared in the coherent state $\ket{\alpha_{\text{c}}}$ of the $\text{HG}_{00}$ spatial mode, with mean photon number $N_{0, {\text{c}}}=|\alpha_{\text{c}}|^2$. At the receiving satellite, balanced homodyne detection is used to measure the $\text{HG}_{10}$ mode component of the received beam to estimate the displacement of the $\text{HG}_{00}$ beam~\cite{SqueezingHomodyneTheory2004}. This scheme is classically optimal in the sense that it attains the SQL for the minimum achievable displacement-estimation uncertainty $\Delta d_{\text{SQL}}$~\cite{TEM10_HomodyneOptimal2006}.

\subsubsection{Quantum-Enhanced Scheme}\label{Sec:QES}
It has been shown in~\cite{QuantumNoise2005} that in order to achieve a precision beyond that dictated by the SQL, adding squeezing to the appropriate transverse mode of the input field is a necessary and sufficient condition. Here, we consider a practical quantum-enhanced scheme that enhances the QFI by adding squeezing to the derivative mode $u_1[d]\propto \partial_d u_0[d]$. 
In this quantum-enhanced scheme, the laser probe beam is a multimode field in which the $\text{HG}_{00}$ spatial mode is prepared in the coherent state $\ket{\alpha_{\text{q}}}$  (with mean photon number $N_{0,\text{q}}=|\alpha_{\text{q}}|^2$), while the $\text{HG}_{10}$ spatial mode\footnote{In the absence of any spatial truncation at the transmitter, in the limit $d\rightarrow 0$, this derivative mode is proportional to the $\text{HG}_{10}$ mode since the properties of Hermite polynomials give $\left.\partial_d u_0(x-d,y,z)\right|_{d\rightarrow0}=-\partial_x u_0(x,y,z)$.}
is prepared in a squeezed-vacuum state with squeezing parameter $r_{\text{in}}$. Note that we assume no correlation between the $\text{HG}_{00}$ and $\text{HG}_{10}$ modes.
At the receiving satellite, balanced homodyne detection is used to measure the relevant quadrature of the $\text{HG}_{10}$-mode component of the received beam, from which the displacement of the transmitted beam is estimated. When the fluctuation of this quadrature is reduced below the vacuum level, the minimum achievable displacement-estimation uncertainty can be reduced below that dictated by the SQL~\cite{TEM10_HomodyneOptimal2006,DetectionNoiseModesOptimal2006,UltimateSensitivity2012}.

\subsection{Quantum Advantage}\label{Sec:QA}
Consider the minimum achievable displacement-estimation uncertainty $\Delta d$, defined as the smallest standard deviation attainable by an unbiased displacement estimator. Intuitively, a smaller $\Delta d$ indicates a higher level of precision and therefore better tracking performance.  
According to the Cram\'er–Rao bound, $\Delta d$ is inversely proportional to the square root of the QFI (denoted by $\mathcal{F}_d$) of the quantum state $\hat{\rho}_{d}$ in which the displacement $d$ is encoded~\cite{AdvQuantumMetrology2011}. As a result, the quantum advantage can be formally defined as
\begin{equation}\label{Eq:QA_General}
    Q=\frac{\Delta d_{\text{SQL}}}{\Delta d_{\text{QE}}}=\sqrt{\frac{\mathcal{F}_{d}^{{\text{QE}}}}{\mathcal{F}_{d}^{{\text{SQL}}}}},
\end{equation}
where the labels QE and SQL denote quantities associated with the quantum-enhanced scheme and the SQL-limited optimal classical scheme considered in this work, respectively.

In the following sections we will perform theoretical analyses to evaluate the QFI. 
It should be noted that although the optimal measurements for both of the investigated schemes are well known and have already been demonstrated (in, e.g.,~\cite{TEM10_HomodyneOptimal2006}), our analyses are inherently measurement-independent since the use of quantum parameter estimation theory and the QFI formalism involves an optimization over all possible measurements.

\section{Theoretical Analyses}

In this section, we apply the methodology introduced by~\cite{GaussianQFI2024} (detailed in Appendix~\ref{Sec:ModeEncodedPE}) to evaluate the QFI in the two considered displacement-estimation schemes and quantify the quantum advantage within the context of inter-satellite QSTT. In particular, the work reported here can be regarded as an extension of the spatial-beam-positioning example in~\cite{GaussianQFI2024} to an inter-satellite deployment scenario. 

\subsection{Spatial-Mode Basis at Receiving Satellite}\label{Sec:DetectionModes}
The framework of~\cite{GaussianQFI2024} allows one to evaluate the QFI for parameters encoded into both the quantum state and the spatial-temporal mode structure of an optical field. The parameter of interest in this work, the transverse displacement $d$, is such a parameter. An infinitesimal variation $\delta d$ in $d$ induces, to first order, modal coupling associated with the derivatives of the initially populated spatial modes (those modes leaving the transmitter), with coupling amplitudes proportional to $\delta d$. The full parameter-sensitive spatial-mode subspace can therefore be constructed from the initially populated spatial modes together with the additional spatial modes generated by their derivatives. These derivative modes identify the additional orthogonal modal directions into which the initially populated modes are coupled under an infinitesimal variation of $d$ and are  required for evaluating the QFI. 

It should be noted that aperture truncation is not considered in~\cite{GaussianQFI2024}; therefore, an additional receiver-side treatment is required in our case. The initially populated spatial modes $\{u_k[d](x,y,z)\}$ describe the field \textit{before} aperture truncation, whereas the parameter-sensitive spatial-mode subspace at the receiver must be constructed from the corresponding aperture-truncated received modes together with the orthogonal components of their derivatives with respect to $d$. The aperture truncation changes the norms (the square roots of self-inner products) of the spatial-mode functions entering the receiver-side parameter-sensitive subspace (before renormalization).

To help the reader, we summarize the mathematical pathway we now embark upon. This pathway will lead us to the main result of this paper, namely Eq.~(\ref{Eq:GaussianQFI_TwoMode}) -- a determination of the QFI on beam displacement within a multimode system that suffers real-world truncation at the receiver. Using this equation, we then determine the quantum advantage of our quantum-enhanced scheme relative to the optimal classical scheme normalized to the same energy input.

Step 1:
We first construct an orthonormal basis for this receiver-side parameter-sensitive subspace and refer to the resulting basis modes as the \textit{detection modes} (denoted by $w_n$ with $n$ being an integer subscript).\footnote{To avoid confusion,  we use $w$ (without a subscript) to denote the beam radius.}
Specifically,  $w_0[d](x,y,z)$ denotes the normalized aperture-truncated version of the displaced HG$_{00}$ mode, $w_1[d](x,y,z)$ denotes the normalized component of the aperture-truncated source derivative mode $A u_1[d]$ that is orthogonal to $w_0[d]$, and $w_2[d](x,y,z)$ denotes the orthonormalized component of $\partial_d w_1[d](x,y,z)$ that is orthogonal to both $w_0[d]$ and $w_1[d]$. [Step 1 is detailed in Sec.~\ref{Appendix:GS} parts~\textit{a-b}.]

Step 2: From the modes derived in the above step we then simplify them for a system at our operating point \(d\rightarrow0\). As we shall see, at the symmetric operating point $d=0$, the parity of the HG modes together with the symmetry of the centered circular aperture gives $\langle A u_0[0], A u_1[0]\rangle=0$. Consequently, in the limit $d \rightarrow 0$, $w_0[d]$ and $w_1[d]$ reduce to the aperture-truncated and renormalized versions of the HG$_{00}$ and HG$_{10}$ modes, respectively. The mode $w_2[d](x,y,z)$, however, is generally not an aperture-truncated HG$_{20}$ mode. [Step 2 is detailed in Sec.~\ref{Appendix:GS} part~\textit{c}.]

Step 3: As we shall see the detection modes derived from the above steps depend on three important quantities -- referred to as mode-related coefficients. Two of these involve inner products of spatial-mode functions and one is the norm of the orthogonal component of a mode derivative. Analytical expressions for these three coefficients are required in order for us to move forward, and in this step these expressions are derived. [Step 3  is detailed  in Sec.~\ref{Appendix:Coeffs}.]

Step 4: 
In this step, the QFI is determined for the optimal classical scheme, in which the probe consists only of a coherent-state component without squeezing. The corresponding QFI determines the quantum Cram\'er-Rao bound, which defines the SQL considered in this work and serves as the classical benchmark.
[Step 4 is detailed in Sec.~\ref{qficlass}.]

Step 5: 
In this step, the QFI for our quantum-enhanced scheme is determined. In this scheme, a squeezed-vacuum component is introduced in the first derivative spatial mode. The main result of this work is derived in this section.
[Step 5 is detailed in Sec.~\ref{qfiQ}.]

Let us commence our derivation.

\subsubsection{Gram-Schmidt Orthonormalization for Detection Modes}\label{Appendix:GS}

Here we give the full construction of the detection modes $w_0[d]$, $w_1[d]$, and $w_2[d]$ via Gram-Schmidt orthonormalization. Note that, when analyzing spatial modes, the longitudinal coordinate $z$ and the Gouy phase are omitted for simplicity, and the transverse spatial coordinates $(x,y)$ are omitted whenever possible.

The displaced Gaussian beam at the receiver is given by
\begin{equation}\label{Eq:DisplacedHG00}
	u_0[d](x,y)
	=
	\frac{1}{\sqrt{\pi w^2}}
	\exp\!\left[
	-\frac{(x-d)^2+y^2}{2w^2}
	\right].
\end{equation}
The normalized derivative mode of Eq.~(\ref{Eq:DisplacedHG00}) is given by
\begin{equation}\label{Eq:u_1_u_0_eta}
	u_1[d]=\frac{\partial_d u_0[d]}{\eta},
\end{equation}
where $\eta=\frac{1}{\sqrt{2}\,w}$. At $d=0$, $u_1[0]$ is the $\text{HG}_{10}$ mode.

\paragraph{Aperture-truncated source-mode functions}\label{Sec:Aperture}

The circular receiving aperture is given by
\begin{equation}\label{Eq:Ap}
	A := A(x,y)
	=
	\begin{cases}
		1, & x^2+y^2 \le a^2,\\
		0, & x^2+y^2>a^2
	\end{cases}.
\end{equation}
As a result, the truncated source-mode functions are given by
\begin{equation}
	f_0[d]:=A\,u_0[d],
	\qquad
	f_1[d]:=A\,u_1[d].
\end{equation}
These are the functions on which the Gram-Schmidt procedure is performed to find the detection modes.

\paragraph{Construction of the detection modes}

The first detection mode $w_0[d]$ is given by the normalization of $f_0[d]$,
\begin{equation}\label{Eq:w0d}
	w_0[d] =\frac{A u_0[d]}{\|A u_0[d]\|} = \frac{f_0[d]}{\sqrt{\kappa_0(d)}},
\end{equation}
where $\kappa_0(d)=\langle f_0[d],f_0[d]\rangle=\|f_0[d]\|^2$ is the squared norm of $f_0[d]$.
As a result, $\langle w_0[d],w_0[d]\rangle=1$. Note that $\langle x,y\rangle$ represents the inner product of the functions $x$ and $y$.

Employing the Gram-Schmidt process, the second detection mode $w_1[d]$ is obtained by orthogonalizing $f_1[d]$ against the first detection mode $w_0[d]$,
\begin{equation}\label{Eq:w1d}
		w_1[d]
		=\frac{g_1[d]}{\sqrt{\tau_1(d)}}=
		\frac{
			f_1[d]- s(d)w_0[d]
		}{
			\|f_1[d]-s(d)w_0[d]\|
		},
\end{equation}
where $s(d)=\langle w_0[d],f_1[d]\rangle$ denotes the overlap between $f_1[d]$ and the first detection mode $w_0[d]$. Such a construction guarantees
$\langle w_1[d],w_1[d]\rangle=1$ and $\langle w_0[d],w_1[d]\rangle=0$.

The third detection mode $w_2[d]$ is given by further performing the Gram-Schmidt process on $h_2[d]=\partial_d w_1[d]$, which is the derivative of the second detection mode $w_1[d]$,against both $w_0[d]$ and $w_1[d]$. Specifically, the third detection mode $w_2[d]$ is given by 
\begin{equation}
		w_2[d]
		=
		\frac{
			g_2[d]
		}{
			\left\|
		g_2[d]
			\right\|
		},
\end{equation}
where the Gram-Schmidt process leads to
\begin{equation}\label{Eq:g2}
g_2[d] = h_2[d]-\langle w_0[d],h_2[d]\rangle\,w_0[d]-\langle w_1[d],h_2[d]\rangle\,w_1[d].
\end{equation}
Since $\langle w_1[d],w_1[d]\rangle=1$, differentiating this identity with respect to $d$ gives
\begin{equation}
	\partial_d \langle w_1[d],w_1[d]\rangle = 0.
\end{equation}
Following the chain rule, we have
\begin{equation}\label{Eq:ChainRulew1}
	\langle \partial_d w_1[d],w_1[d]\rangle+\langle w_1[d],\partial_d w_1[d]\rangle=0.
\end{equation}

Since $w_1[d]$ and $\partial_d w_1[d]$ are taken to be real spatial-mode functions, Eq.~(\ref{Eq:ChainRulew1}) indicates that
\begin{equation}
	\langle \partial_d w_1[d], w_1[d]\rangle=\langle w_1[d],\partial_d w_1[d]\rangle=0.
\end{equation}
Therefore, the $\langle w_1[d],h_2[d]\rangle$ term in Eq.~(\ref{Eq:g2}) vanishes, and the expression for $w_2[d]$ reduces to
\begin{equation}\label{Eq:w2d}
		w_2[d]
		=
		\frac{
			\partial_d w_1[d]-\xi_A(d)w_0[d]
		}{
			\zeta_A(d)
		},
\end{equation}
where
\begin{equation}\label{Eq:xid}
	\xi_A(d):=\langle w_0[d],\partial_d w_1[d]\rangle,
\end{equation}
and
\begin{equation}\label{Eq:zetad}
	\zeta_A(d):=
	\left\|
	\partial_d w_1[d]-\xi_A(d)w_0[d]
	\right\|.
\end{equation}

\paragraph{Detection modes at $d\rightarrow 0$}
At $d=0$, it is clear that $u_0[0]$ is even in $x$, and the first derivative mode $u_1[0]$ is odd in $x$. The aperture function $A(x,y)$ is even in $x$.
As a result, $f_0[0]=A u_0[0]$ ($f_1[0]=A u_1[0]$) is even (odd), and $s(0)=\langle w_0[0],f_1[0]\rangle=0$. This leads to the following simplified expressions for the first two detection modes,
\begin{equation}\label{Eq:w01d}
\begin{aligned}
	w_0[0] &:= w_0[d]\big{|}_{d \rightarrow 0}=\frac{A u_0[0]}{\sqrt{\kappa_0}},\\
	w_1[0] &:= w_1[d]\big{|}_{d \rightarrow 0}=\frac{A u_1[0]}{\sqrt{\kappa_1}}.
\end{aligned}
\end{equation}
In Eq.~(\ref{Eq:w01d}), we have defined
\begin{equation}\label{Eq:k0_and_k1}
\begin{aligned}
	\kappa_0&:=\kappa_0(d)\big{|}_{d \rightarrow 0}\!\!=\|A u_0[0]\|^2=1-e^{-A_0}, \\
	\kappa_1&:=\kappa_1(d)\big{|}_{d \rightarrow 0}\!\!=\|A u_1[0]\|^2=1-\left(A_0+1\right)e^{-A_0},
\end{aligned}
\end{equation}
where  $A_0:={a^2}/{w^2}$.

The coefficient $\eta_A(d)$ is an important factor relating $\partial_d w_0[d]$ to the second detection mode $w_1[d]$. In the present mode construction, $\partial_d w_0[d]$ is proportional to the Gram-Schmidt-orthonormalized mode $w_1[d]$. We therefore define
\begin{equation}\label{Eq:etaA_d}
     \eta_A(d)=\left\langle \partial_d w_0[d], w_1[d]\right\rangle,
\end{equation}
so that
\begin{equation}\label{Eq:w1_eta_d}
     w_1[d]=\frac{\partial_d w_0[d]}{\eta_A(d)}.
\end{equation}
At $d\rightarrow 0$, we define 
\begin{equation}\label{Eq:etaA}
\eta_A:=\eta_A(0)=\left\langle \partial_d w_0[d]\big{|}_{d \rightarrow 0}, w_1[d]\big{|}_{d \rightarrow 0}\right\rangle.
\end{equation}

The coefficient $s(d)$ in Eq.~(\ref{Eq:w1d}) is given by
\begin{equation}\label{Eq:sd}
	s(d)=\langle w_0[d],f_1[d]\rangle=\frac{\langle f_0[d],f_1[d]\rangle}{\sqrt{\kappa_0(d)}}.
\end{equation}
Differentiating $\kappa_0(d)=\langle A u_0[d],A u_0[d]\rangle$, we have
\begin{equation}\label{Eq:DiffKappa0d}
\begin{aligned}
\partial_{d}\kappa_0(d)
	&=
	\langle A \partial_d u_0[d],A u_0[d]\rangle+\langle A u_0[d],A \partial_d u_0[d]\rangle\\
	&=
	2\langle A u_0[d],A \partial_d u_0[d]\rangle,
\end{aligned}
\end{equation}
since the inner product is real. Using \(\partial_d u_0[d]=\eta u_1[d]\) (Eq.~(\ref{Eq:u_1_u_0_eta})), Eq.~(\ref{Eq:DiffKappa0d}) gives
\begin{equation}
	\langle f_0[d],f_1[d]\rangle=\frac{\partial_d \kappa_0(d)}{2\eta}.
\end{equation}

Using \(\eta=1/(\sqrt{2}w)\), we can rewrite $s(d)$ in Eq.~(\ref{Eq:sd}) as
\begin{equation}\label{Eq:sd1}
	s(d)=\frac{w}{\sqrt{2}}\frac{\partial_d \kappa_0(d)}{\sqrt{\kappa_0(d)}}.
\end{equation}
Differentiating  $s(d)$ in Eq.~(\ref{Eq:sd1}), we have
\begin{equation}\label{Eq:ddsd}
\partial_d s(d)
	=
	\frac{w}{\sqrt{2}}
	\left(
	\frac{\partial_d^2 \kappa_0(d)}{\sqrt{\kappa_0(d)}}
	-
	\frac{(\partial_d \kappa_0(d))^2}{2\kappa_0^{3/2}(d)}
	\right).
\end{equation}
At $d\rightarrow 0$, recalling the symmetry $\partial_d \kappa_0(d)\big{|}_{d \rightarrow 0}=0$, we have
\begin{equation}\label{Eq:dds0}
	\partial_d s(d)\big{|}_{d \rightarrow 0}=\frac{w}{\sqrt{2}}\frac{\partial^2_d\kappa_0(d)\big{|}_{d \rightarrow 0}}{\sqrt{\kappa_0}}.
\end{equation}
Since 
\begin{equation}\label{Eq:k0''}
	\partial^2_{d}\kappa_0(d)\big{|}_{d \rightarrow 0}\!\!=-\frac{2A_0 e^{-A_0}}{w^2},
\end{equation}
at $d\rightarrow 0$, $\partial_d s(d)\big{|}_{d \rightarrow 0}$ given by Eq.~(\ref{Eq:ddsd}) is found to be
\begin{equation}\label{Eq:dds0_exp}
		\partial_d s(d)\big{|}_{d \rightarrow 0}=-\frac{\sqrt{2}A_0 e^{-A_0}}{w\sqrt{1-e^{-A_0}}}.
\end{equation}

At $d\rightarrow 0$, the third detection mode $w_2$ defined by Eq.~(\ref{Eq:w2d}) becomes 
\begin{equation}\label{Eq:w20}
    w_2[d]\big{|}_{d \rightarrow 0}=\frac{\partial_d w_1[d]\big{|}_{d \rightarrow 0}-\xi_A w_0[d]\big{|}_{d \rightarrow 0}}{\zeta_A},
\end{equation}
where we have defined $\xi_A:=\xi_A(0)$ and $\zeta_A:=\zeta_A(0)$. Equivalently, Eq.~(\ref{Eq:w20}) can be rearranged as
\begin{equation}\label{Eq:dw1_decomp}
    \partial_d w_1[d]\big{|}_{d\rightarrow 0}
    =
    \xi_A w_0[d]\big{|}_{d\rightarrow 0}
    +
    \zeta_A w_2[d]\big{|}_{d\rightarrow 0}.
\end{equation}
Thus, at the symmetric operating point, \(\partial_d w_1\) (the derivative of $w_1$) decomposes into a component along \(w_0\) and an orthogonal component along \(w_2\), with coefficients \(\xi_A\) and \(\zeta_A\), respectively. To obtain an explicit expression for the third detection mode $w_2$ in terms of the aperture-truncated functions and the quantities introduced in the receiver-side Gram-Schmidt construction, we next evaluate \(\partial_d w_1[d]\big|_{d\rightarrow0}\) in Eq.~(\ref{Eq:w20}).
For this purpose, we differentiate $w_1[d]=\frac{g_1[d]}{\sqrt{\tau_1(d)}}$ in Eq.~(\ref{Eq:w1d}) and have 
\begin{equation}
	\partial_d w_1[d]=\frac{\partial_d g_1[d]}{\sqrt{\tau_1(d)}}-\frac{\partial_d \tau_1(d)}{2\tau_1(d)}w_1[d].
\end{equation}
Differentiating $g_1[d]=f_1[d]-s(d)w_0[d]$, we have
\begin{equation}
	\partial_d g_1[d]=\partial_d f_1[d]-\partial_d s(d)w_0[d]-s(d)\partial_d w_0[d].
\end{equation}
At \(d\rightarrow 0\), since \(s(0)=0\), the above equation reduces to
\begin{equation}
	\partial_d g_1[d]\big{|}_{d \rightarrow 0} = \partial_d f_1[d]\big{|}_{d \rightarrow 0}-\partial_d s(d)\big{|}_{d \rightarrow 0} w_0[0].
\end{equation}

We now derive $\partial_d \tau_1(d)$ at $d \rightarrow 0$. Since $\tau_1(d)=\|g_1[d]\|^2$ and $g_1[d]=f_1[d]-s(d)w_0[d]$ with $w_0[d]$ being normalized and $g_1[d]\perp w_0[d]$, we have
\begin{equation}
	\tau_1(d)=\|f_1[d]\|^2-|s(d)|^2=\kappa_1(d)-|s(d)|^2.
\end{equation}
Differentiating $\tau_1(d)$ in the above equation, we have
\begin{equation}
	\partial_d \tau_1(d)=\partial_d \kappa_1(d)-2s(d)\partial_d s(d).
\end{equation}
Recalling $s(d)=0$ at $d \rightarrow 0$, we can see that
\begin{equation}
	\partial_d \tau_1(d) \big{|}_{d \rightarrow 0}\!\!=\partial_d \kappa_1(d)\big{|}_{d \rightarrow 0}\!\!=0,
\end{equation}
where we have used the property $\partial_d \kappa_1(d)\big{|}_{d \rightarrow 0}\!\!\!\!=0$ since $\kappa_1(d)$ is an even function of $d$.
Since \(\tau_1(0)=\kappa_1\) and $\partial_d\tau_1(d)\big|_{d\rightarrow0}=0$, the first term in the numerator of Eq.~(\ref{Eq:w20}) can be expressed as
\begin{equation}\label{Eq:ddw10}
	\partial_d w_1[d]\big{|}_{d \rightarrow 0}\!\!=\frac{\partial_d f_1[d]\big{|}_{d \rightarrow 0}-\partial_d s(d)\big{|}_{d \rightarrow 0} w_0[d]\big{|}_{d \rightarrow 0}}{\sqrt{\kappa_1}}.
\end{equation}
Substituting the expression for $\partial_d w_1[d]\big{|}_{d \rightarrow 0}$ in Eq.~(\ref{Eq:ddw10}), we can express $w_2[d]\big{|}_{d \rightarrow 0}$ in Eq.~(\ref{Eq:w20}) as
\begin{align}
	w_2[d]\big{|}_{d \rightarrow 0} = \frac{
	\frac{\partial_d f_1[d]\big{|}_{d \rightarrow 0}}{\sqrt{\kappa_1}}
	-\left(
	\frac{\partial_d s(d)\big{|}_{d \rightarrow 0}}{\sqrt{\kappa_1}}+\xi_A
	\right)w_0[d]\big{|}_{d \rightarrow 0}}{\zeta_A},
\end{align}
where $\partial_d s(d)\big{|}_{d \rightarrow 0}$ is given by Eq.~(\ref{Eq:dds0_exp}). 
This expression provides an explicit receiver-side representation of the third detection mode \(w_2\) in terms of the derivative of the aperture-truncated mode \(f_1\), the Gram-Schmidt overlap quantity \(s(d)\), and the coefficients \(\xi_A\) and \(\zeta_A\).
The full derivation of the important mode-related coefficients $\eta_{\text{A}}$, $\xi_{\text{A}}$, and $\zeta_{\text{A}}$ is provided in Sec.~\ref{Appendix:Coeffs}.

\subsubsection{Analytical Expressions for Important Mode-Related Coefficients}\label{Appendix:Coeffs}

Here we provide the full derivation of the important mode-related coefficients $\eta_{\text{A}}$, $\xi_{\text{A}}$, and $\zeta_{\text{A}}$ defined in Sec.~\ref{Appendix:GS}.

\paragraph{Analytical Expression for $\eta_A$ in Eq.~(\ref{Eq:etaA}) at $d\rightarrow 0$}\label{Appendix:etaA}

Differentiating $w_0[d]=\frac{f_0[d]}{\sqrt{\kappa_0(d)}}$ in Eq.~(\ref{Eq:w0d}), we have
\begin{equation}
	\partial_d w_0[d]=\frac{\partial_d f_0[d]}{\sqrt{\kappa_0(d)}}-\frac{\partial_d\kappa_0(d)}{2\kappa_0(d)}w_0[d].
\end{equation}
At $d=0$, we have $\partial_d \kappa_0(d)=0$, and the above equation is reduced to
\begin{equation}
	\partial_d w_0[d]\big{|}_{d \rightarrow 0} =\frac{\partial_d f_0[d]\big{|}_{d \rightarrow 0}}{\sqrt{\kappa_0}}.
\end{equation}

Since $\partial_d f_0[d]\big{|}_{d \rightarrow 0}\!\!\!\!=A \partial_d u_0[d]\big{|}_{d \rightarrow 0}$ and $u_1[d]\big{|}_{d \rightarrow 0}=\frac{\partial_d u_0[d]\big{|}_{d \rightarrow 0}}{\eta}$, we have
\begin{equation}
	\partial_d w_0[d]\big{|}_{d \rightarrow 0}=\eta\frac{A u_1[d]\big{|}_{d \rightarrow 0}}{\sqrt{\kappa_0}}=\eta\sqrt{\frac{\kappa_1}{\kappa_0}}\,w_1[d]\big{|}_{d \rightarrow 0}.
\end{equation}
Therefore, at $d\rightarrow 0$, $\eta_A$ is given by
\begin{equation}\label{Eq:etaA_final}
	\eta_A=\left\langle \partial_d w_0[d]\big{|}_{d \rightarrow 0},w_1[d]\big{|}_{d \rightarrow 0}\right\rangle=\eta\sqrt{\frac{\kappa_1}{\kappa_0}}=\sqrt{\frac{\kappa_1}{2w^2\kappa_0}}.
\end{equation}

\paragraph{Analytical Expression for $\xi_A$ in Eq.~(\ref{Eq:xid}) at $d\rightarrow 0$}
Since $\langle w_0[d],w_1[d]\rangle=0$, we have
\begin{equation}
	\langle \partial_d w_0[d],w_1[d]\rangle+\langle w_0[d],\partial_d w_1[d]\rangle=0.
\end{equation}
At $d \rightarrow 0$, the above equation becomes
\begin{equation}\label{Eq:etaA_and_xiA}
	\langle \partial_d w_0[d]\big{|}_{d \rightarrow 0},w_1[d]\big{|}_{d \rightarrow 0}\rangle+\langle w_0[d]\big{|}_{d \rightarrow 0},\partial_d w_1[d]\big{|}_{d \rightarrow 0}\rangle=0.
\end{equation}
In Eq.~(\ref{Eq:etaA_and_xiA}), the second term is exactly \(\xi_A\), while the first term can be rewritten as
\begin{equation}
	\langle \partial_d w_0[d]\big{|}_{d \rightarrow 0},w_1[d]\big{|}_{d \rightarrow 0}\rangle
	=
	\langle \eta_A w_1[0],w_1[0]\rangle
	=
	\eta_A.
\end{equation}
As a result, Eq.~(\ref{Eq:etaA_and_xiA}) can be re-expressed as
\begin{equation}
	\eta_A+\xi_A=0,
\end{equation}
leading to
\begin{equation}\label{Eq:xiA_Final}
		\xi_A=-\eta_A.
\end{equation}

\paragraph{Analytical Expression for $\zeta_A$ in Eq.~(\ref{Eq:zetad}) at $d\rightarrow 0$}

Differentiating $w_1[d]=\frac{\partial_d w_0[d]}{\eta_A(d)}$ in Eq.~(\ref{Eq:w1_eta_d}), we have 
\begin{equation}
	\partial_d w_1[d]=\frac{\partial^2_d w_0[d]}{\eta_A(d)}-\frac{\partial_d \eta_A(d)}{\eta_A^2(d)}\partial_d w_0[d].
\end{equation}
Noting that by symmetry we have $\partial_d \eta_A(d)\big{|}_{d \rightarrow 0}=0$ since $\eta_A(d)$ is an even function of $d$, at $d \rightarrow 0$ the above equation can be reduced to
\begin{equation}
	\partial_d w_1[d]\big{|}_{d \rightarrow 0}=\frac{\partial^2_d w_0[d]\big{|}_{d \rightarrow 0}}{\eta_A}.
\end{equation}
Therefore,
\begin{equation}\label{Eq:zeta2}
	\zeta_A^2=\|\partial_d w_1[d]\big{|}_{d \rightarrow 0}\|^2-\xi_A^2=\frac{\|\partial^2_d w_0[d]\big{|}_{d \rightarrow 0}\|^2}{\eta_A^2}-\eta_A^2.
\end{equation}

 Next, we derive $\partial^2_d w_0[d]\big{|}_{d \rightarrow 0}$ in Eq.~(\ref{Eq:zeta2}).
Differentiating $w_0[d]=\frac{f_0[d]}{\sqrt{\kappa_0(d)}}$ twice, at $d\rightarrow 0$ we have
\begin{equation}\label{Eq:dd2w00}
	\partial^2_d w_0[d]\big{|}_{d \rightarrow 0}=\frac{\partial^2_d f_0[d]\big{|}_{d \rightarrow 0}}{\sqrt{\kappa_0}}-\frac{\partial^2_d\kappa_0(d)\big{|}_{d \rightarrow 0}}{2\kappa_0}w_0[d]\big{|}_{d \rightarrow 0},
\end{equation}
where we have again used $\partial_d \kappa_0(d)\big{|}_{d \rightarrow 0}=0$.
Since 
\begin{equation}
	\partial^2_d u_0[d]\big{|}_{d \rightarrow 0}=\left(\frac{x^2}{w^4}-\frac{1}{w^2}\right)u_0[d]\big{|}_{d \rightarrow 0},
\end{equation}
we have 
\begin{equation}\label{Eq:dd2f0d}
	\partial_d^2 f_0[d]\big{|}_{d \rightarrow 0} =A \partial_d^2 u_0[d]\big{|}_{d \rightarrow 0}
	=
	\left(\frac{x^2}{w^4}-\frac{1}{w^2}\right)A u_0[d]\big{|}_{d \rightarrow 0}.
\end{equation}
Substituting $A u_0[0]=\sqrt{\kappa_0}\,w_0[0]$ into Eq.~(\ref{Eq:dd2f0d}) and substituting Eq.~(\ref{Eq:dd2f0d}) into Eq.~(\ref{Eq:dd2w00}) we have
\begin{equation}
\begin{aligned}
	\partial_d^2 w_0[d]\big{|}_{d \rightarrow 0} &=
	\left(
	\frac{x^2}{w^4}
	-\frac{1}{w^2}
	-\frac{\partial_d^2 \kappa_0(d)\big{|}_{d \rightarrow 0}}{2\kappa_0}
	\right)w_0[0]\\
    &=\left(\frac{x^2}{w^4}-\frac{\beta}{w^2}\right)w_0[0],
\end{aligned}
\end{equation}
where $\beta:=1-\frac{A_0E}{1-E}$ with $E=e^{-A_0}$ and $A_0=\frac{a^2}{w^2}$. In the above equation we have used Eq.~(\ref{Eq:k0''}) for $\partial_d^2 \kappa_0(d)\big{|}_{d \rightarrow 0}$.

The squared norm of $\partial_d^2 w_0[d]\big{|}_{d \rightarrow 0}$ in Eq.~(\ref{Eq:zeta2}) can  then be given by
\begin{equation}\label{Eq:NormOfSecondDerivative}
    \begin{aligned}
	\left\|\partial_d^2 w_0[d]\big{|}_{d \rightarrow 0} \right\|^2
	&=
	\int\!\!\int
	\left(\frac{x^2}{w^4}-\frac{\beta}{w^2}\right)^2 w_0^2[0](x,y)\,\mathrm{d}x\mathrm{d}y\\
	&=
	\frac{\langle x^4\rangle}{w^8}
	-\frac{2\beta\langle x^2\rangle}{w^6}
	+\frac{\beta^2}{w^4}\\
    &= 
    \frac{1}{w^4}
	\left(
	\frac{3}{8}\mu_4-\beta\mu_2+\beta^2
	\right),
\end{aligned}
\end{equation}
where $\mu_2=\frac{1-(A_0+1)E}{1-E}$ and $\mu_4=\frac{2-(A_0^2+2A_0+2)E}{1-E}$ are dimensionless factors determined from the moments evaluated with respect to the normalized aperture-truncated intensity profile $w_0^2[0](x,y)$.

Finally, substituting $\|\partial_d^2 w_0[d]\big{|}_{d \rightarrow 0} \|^2$ (given by Eq.~(\ref{Eq:NormOfSecondDerivative})) and $\eta_A^2$ (from Eq.~(\ref{Eq:etaA_final})) into Eq.~(\ref{Eq:zeta2}), we obtain
\begin{equation}\label{Eq:zetaA_Final}
		\zeta_A
		=
		\sqrt{\frac{1}{w^2}
		\left[
		\frac{3\mu_4}{4\mu_2}
		-2\beta
		+\frac{2\beta^2}{\mu_2}
		-\frac{\mu_2}{2}
		\right]}.
\end{equation}

\subsection{QFI Analyses}
We now determine the Fisher information for both the classical scheme and the quantum-enhanced scheme. This allows us to determine the presence of any quantum advantage from a \textit{fundamental} perspective, irrespective of the detailed measurement scheme used to determine the beam displacement in each scheme.

\subsubsection{QFI for Optimal Classical Scheme} \label{qficlass}
In the optimal classical scheme described in Sec.~\ref{Sec:OptCS}, the laser probe beam is a single-mode field prepared in the coherent state $\ket{\alpha_{\text{c}}}$ of the $\text{HG}_{00}$ spatial mode, with mean photon number $N_{0, {\text{c}}}=|\alpha_{\text{c}}|^2$. 
In this scheme, the displacement vector of the corresponding single-mode Gaussian state at the receiving satellite (expressed in the receiver-side detection mode $w_0$) can be written as $\bar{\mathbf{x}}_0^{\top}=[q_0,p_0]=[2\sqrt{\kappa_0} \Re{\alpha_{\text{c}}},2\sqrt{\kappa_0}\Im{\alpha_{\text{c}}}]$, where $\kappa_0=1-e^{-{a^2}/{w^2}}$ is the transmissivity associated with the $\text{HG}_{00}$ source mode (derived in Eq.~(\ref{Eq:k0_and_k1})). The covariance matrix of the received Gaussian state (expressed in the receiver-side detection mode $w_0$) is $V_0=\mathbbm{1}_2$.

After determining the coefficients in Eq.~(\ref{Eq:CoeffsForDMatrixGeneral}) using our orthonormalization procedure detailed in Sec.~\ref{Appendix:GS}, we have
\begin{equation}
    D_n=0,\quad D_{\partial}=\eta_A \mathbbm{1}_2,
\end{equation}
where $\eta_A$ is defined in Eq.~(\ref{Eq:etaA_d}) with analytical expression given by Eq.~(\ref{Eq:etaA_final}). 

The displacement-vector contribution to the QFI (see Eq.~(\ref{Eq:DispQFI_General})) is then simplified to
\begin{equation}
    F_{\bar{\mathbf{x}}}=\left(\partial_d \bar{\mathbf{x}}_0\right)^{\top} V_0^{-1}\left(\partial_d \bar{\mathbf{x}}_0\right)+\eta_A^2\left\|\bar{\mathbf{x}}_0\right\|^2.
\end{equation}

The covariance-matrix contribution to the QFI (see Eq.~(\ref{Eq:CovQFI_General})) can be simplified to
\begin{equation}
    F_\sigma=\frac{1}{2} \sum_{l=0}^3\left(\frac{a_l^2}{\nu_0^2-(-1)^l}+\frac{b_l^2}{\nu_0-(-1)^l}\right),
\end{equation}
where
\begin{equation}\label{Eq:Coeffs_a_b_in_SingleMode}
\begin{aligned}
& a_l^2=\operatorname{tr}\left[A_l S_0^{-1}\left(\partial_d V_0\right)\left(S_0^{\top}\right)^{-1}\right]^2, \\
& b_l^2=\eta_A^2 \operatorname{tr}\left[A_l\left(V_0-\mathbbm{1}_2\right)\left(S_0^{\top}\right)^{-1}\right]^2+\eta_A^2 \operatorname{tr}\left[A_l S_0^{-1}\left(V_0-\mathbbm{1}_2\right)\right]^2.
\end{aligned}
\end{equation}
In Eq.~(\ref{Eq:Coeffs_a_b_in_SingleMode}), we have
\begin{equation}\label{Eq:Simplified_A_Matrices}
    A_0=i \sigma_y / \sqrt{2}, A_1=\sigma_z / \sqrt{2}, A_2=\mathbbm{1}_2 / \sqrt{2}, A_3=\sigma_x / \sqrt{2},
\end{equation}
where $\sigma_x$, $\sigma_y$, and $\sigma_z$ are the Pauli $x$, $y$, and $z$ matrices, respectively.

Under the assumption in Sec.~\ref{Sec:Assumptions} that the mean field (represented by the displacement vector in phase space) does not depend explicitly on the beam displacement $d$, the QFI for the optimal classical scheme is given by
\begin{equation}\label{Eq:GaussianQFI_SingleMode}
\mathcal{F}_{d}^{\text{SQL}}=F_{\bar{\mathbf{x}}} + F_\sigma = 4 \kappa_0 \eta_A^2 N_{0,\text{c}},
\end{equation}
where we have used $\partial_d \bar{\mathbf{x}}_0=0$ in the displacement-vector contribution $F_{\bar{\mathbf{x}}}$ and found the covariance-matrix contribution to be $F_\sigma=0$. We recall that the term $N_{0,\text{c}}=|\alpha_{\text{c}}|^2$ in Eq.~(\ref{Eq:GaussianQFI_SingleMode}) quantifies the mean photon number in the transmitted laser probe beam, which is entirely in the $\text{HG}_{00}$ mode.

\subsubsection{QFI for Quantum-Enhanced Scheme}\label{qfiQ}
In the quantum-enhanced scheme described in Sec.~\ref{Sec:QES}, the laser probe beam is a two-mode field in which the $\text{HG}_{00}$ spatial mode is prepared in the coherent state $\ket{\alpha_{\text{q}}}$ (with mean photon number $N_{0,\text{q}}=|\alpha_{\text{q}}|^2$), while the $\text{HG}_{10}$ spatial mode is prepared in a squeezed-vacuum state with squeezing parameter $r_{\text{in}}$. 
In this scheme, the displacement vector of the corresponding two-mode Gaussian state at the receiving satellite (expressed in the receiver-side detection-mode basis $\{w_0, w_1\}$) can be written as $\bar{\mathbf{x}}_2^{\top}=[\bar{\mathbf{x}}_0^{\top},\bar{\mathbf{x}}_1^{\top}]$, where $\bar{\mathbf{x}}_0^{\top}=[q_0,p_0]=[2\sqrt{\kappa_0} \Re{\alpha_{\text{q}}},2\sqrt{\kappa_0}\Im{\alpha_{\text{q}}}]$ and $\bar{\mathbf{x}}_1^{\top}=[q_1,p_1]=[0,0]$.

Since the initial  laser probe beam has no correlations between the two populated modes, the initial covariance matrix of the corresponding two-mode Gaussian state (expressed in the source-mode basis $\{u_0, u_1\}$) is block diagonal. The covariance matrix remains block diagonal after beam propagation and aperture truncation since, in the limit $d\rightarrow 0$, the circular-aperture truncation preserves the orthogonality of the $\text{HG}_{00}$ and $\text{HG}_{10}$ modal components. As a result, the covariance matrix of the received two-mode Gaussian state (expressed in the receiver-side detection-mode basis $\{w_0, w_1\}$) is given by
\begin{equation}\label{Eq:TwoModeStateCM}
    V_2=\left[\begin{array}{cc}
V_0 & 0 \\
0 & V_1
\end{array}\right],
\end{equation}
where $V_0$ denotes the covariance matrix of the first mode (which remains that of a coherent state despite attenuation of its mean amplitude) and $V_1$ denotes the covariance matrix of the second mode (which corresponds to the input squeezed-vacuum state after mixing with vacuum through an effective pure-loss channel). Specifically, $V_0$ and $V_1$ are given by
\begin{equation}\label{Eq:CovBlockQE}
    V_0=\mathbbm{1}_2, \quad V_1=\left[\begin{array}{cc}
\kappa_1 e^{-2 r_{\text{in}}} + 1-\kappa_1 & 0 \\
0 & \kappa_1 e^{2 r_{\text{in}}} + 1-\kappa_1
\end{array}\right],
\end{equation}
where $\kappa_1=1-\left({a^2}/{w^2}+1\right)e^{-{a^2}/{w^2}}$ is the transmissivity associated with the $\text{HG}_{10}$ source mode (defined in Eq.~(\ref{Eq:k0_and_k1})).
Since the QFI is optimized over all possible measurements and the optical phase reference can be chosen freely, we choose the coherent amplitude $\alpha_{\text{q}}$ to be aligned with the squeezed quadrature of the derivative mode; equivalently, $q_0=2\sqrt{\kappa_0}|\alpha_{\text{q}}|$ and $p_0=0$ in the quadrature basis used in Eq.~(\ref{Eq:CovBlockQE}) and in the following analysis.

After determining the coefficients in Eq.~(\ref{Eq:CoeffsForDMatrixGeneral}) using our orthonormalization procedure detailed in Sec.~\ref{Appendix:GS}, we have
\begin{equation}
D_n=\left[\begin{array}{cc}
0 & \eta_A \mathbbm{1}_2 \\
\xi_A \mathbbm{1}_2 & 0
\end{array}\right], \quad D_{\partial}=\left[\begin{array}{c}  0 \\ \zeta_A \mathbbm{1}_2\end{array}\right],
\end{equation}
where $\eta_A$, $\xi_A$, and $\zeta_A$ are defined in Eq.~(\ref{Eq:etaA_d}), Eq.~(\ref{Eq:xid}), and Eq.~(\ref{Eq:zetad}), respectively. The analytical expressions for these coefficients at $d\rightarrow 0$ are given by Eq.~(\ref{Eq:etaA_final}), Eq.~(\ref{Eq:xiA_Final}), and Eq.~(\ref{Eq:zetaA_Final}), respectively.

The displacement-vector contribution to the QFI (see Eq.~(\ref{Eq:DispQFI_General})) is then simplified to
\begin{equation}\label{Eq:FxTwoModeGeneral}
F_{\bar{\mathbf{x}}}=\mathbf{y}^{\top} V_2^{-1} \mathbf{y}=\left[\partial_d \bar{\mathbf{x}}_0^{\top}, \eta_A \bar{\mathbf{x}}_0^{\top}\right] V_2^{-1}
\left[\begin{array}{c} \partial_d \bar{\mathbf{x}}_0 \\ \eta_A \bar{\mathbf{x}}_0 \end{array}\right].
\end{equation}
According to~\cite{UltimateSensitivity2012}, Eq.~(\ref{Eq:FxTwoModeGeneral}) can be rewritten as
\begin{equation}
    F_{\bar{\mathbf{x}}}=\frac{\|\mathbf{y}\|^2}{\left(V_d\right)_{[1,1]}},
\end{equation}
where $\left(V_d\right)_{[1,1]}$ is the variance of the $q$ quadrature of the following mode:
\begin{equation}
\begin{aligned}
    v_0[d](x,y,z)=&\frac{\left(\partial_d q_0+i \partial_d p_0\right)}{\|\mathbf{y}\|} w_0[d](x,y,z)\\
    &\quad\quad\quad\quad\quad\quad+\frac{\eta_A\left(q_0+i p_0\right)}{\|\mathbf{y}\|} w_1[d](x,y,z),
\end{aligned}
\end{equation}
where the first term vanishes since $\partial_d q_0=\partial_d p_0 = 0$ under our assumption.

To evaluate the covariance-matrix contribution to the QFI, we first perform the Williamson decomposition of $V_0$ and $V_1$ (the covariance matrices defined in Eq.~(\ref{Eq:TwoModeStateCM})), finding their corresponding symplectic matrices to be
\begin{equation}\label{Eq:SymMatQE}
\Scale[0.8]{
\begin{aligned}
   S_0 &= \left[
\begin{array}{cc}
 1 & 0 \\
 0 & 1 \\
\end{array}
\right],\\
S_1&=\left(
\begin{array}{cc}
 \sqrt{\frac{{\kappa_1} e^{-2 r_{\text{in}}}-{\kappa_1}+1}{\sqrt{\cosh ^2(r_{\text{in}})-(1-2 {\kappa_1})^2 \sinh ^2(r_{\text{in}})}}} & 0 \\
 0 & \sqrt{\frac{{\kappa_1} e^{2 r_{\text{in}}}-{\kappa_1}+1}{\sqrt{\cosh ^2(r_{\text{in}})-(1-2 {\kappa_1})^2 \sinh ^2(r_{\text{in}})}}} \\
\end{array}
\right),
\end{aligned}}
\end{equation}
with the corresponding symplectic eigenvalues
\begin{equation}\label{Eq:SymEigQE}
\begin{aligned}
    \nu_0 &=1 \quad(\text{for}\,\,V_0), \\
    \nu_1 &=\sqrt{\cosh ^2(r_{\text{in}})-(1-2 \kappa_1)^2 \sinh ^2(r_{\text{in}})} \quad(\text{for}\,\,V_1).
\end{aligned}
\end{equation}

The covariance-matrix contribution to the QFI (see Eq.~(\ref{Eq:CovQFI_General})) can be simplified to
\begin{equation}\label{Eq:CovQFI_Enhanced}
    F_\sigma=\frac{1}{2} \sum_{l=0}^3\left(\frac{a_l^2}{\nu_0^2-(-1)^l}+\frac{{b}_l^2}{\nu_0 \nu_1-(-1)^l}+\frac{d_l^2}{\nu_1-(-1)^l}\right),
\end{equation}
where
\begin{equation}\label{Coeffs_abd_QE}
\Scale[0.8]{
\begin{aligned}
a_l^2= & \operatorname{tr}\left[A_l S_0^{-1}\left(\partial_{d} V_0\right)\left(S_0^{\top}\right)^{-1}\right]^2, \\
{b}_l^2= & \left(\eta_A \operatorname{tr}\left[A_l S_1^{-1}\left(V_0-\mathbbm{1}_2\right)\left(S_0^{\top}\right)^{-1}\right]+\xi_A \operatorname{tr}\left[A_l S_1^{-1}\left(V_1-\mathbbm{1}_2\right)\left(S_0^{\top}\right)^{-1}\right]\right)^2\\
& +\left(\eta_A \operatorname{tr}\left[A_l S_0^{-1}\left(V_0-\mathbbm{1}_2\right)\left(S_1^{-1}\right)^{\top}\right]+\xi_A \operatorname{tr}\left[A_l S_0^{-1}\left(V_1-\mathbbm{1}_2\right)\left(S_1^{-1}\right)^{\top}\right]\right)^2,\\
d_l^2= & \zeta_A^2 \operatorname{tr}\left[A_l S_1^{-1}\left(V_1-\mathbbm{1}_2\right)\right]^2+\zeta_A^2 \operatorname{tr}\left[A_l\left(V_1-\mathbbm{1}_2\right)\left(S_1^{-1}\right)^{\top}\right]^2.
\end{aligned}}
\end{equation}
Substituting the covariance matrices in Eq.~(\ref{Eq:CovBlockQE}), their corresponding symplectic matrices in Eq.~(\ref{Eq:SymMatQE}), and their corresponding symplectic eigenvalues in Eq.~(\ref{Eq:SymEigQE}) into Eq.~(\ref{Coeffs_abd_QE}), we find the non-zero coefficients in Eq.~(\ref{Coeffs_abd_QE}) to be
\begin{equation}\label{Eq:NonZeroCoeffs_QE}
\Scale[0.75]{
\begin{aligned}
b_1^2 &= \frac{\kappa_1^2 e^{-4 r_{\text{in}}} \left(e^{2 r_{\text{in}}}-1\right)^2 \xi_A^2 \left(e^{2 r_{\text{in}}} \sqrt{\kappa_1 \left(e^{-2 r_{\text{in}}}-1\right)+1}+\sqrt{\kappa_1 \left(e^{2 r_{\text{in}}}-1\right)+1}\right)^2}{\sqrt{\cosh ^2(r_{\text{in}})-(1-2 \kappa_1)^2 \sinh ^2(r_{\text{in}})}},\\
b_2^2 &=\frac{\kappa_1^2 e^{-4 r_{\text{in}}} \left(e^{2 r_{\text{in}}}-1\right)^2 \xi_A^2 \left(\sqrt{\kappa_1 \left(e^{2 r_{\text{in}}}-1\right)+1}-e^{2 r_{\text{in}}} \sqrt{\kappa_1 \left(e^{-2 r_{\text{in}}}-1\right)+1}\right)^2}{\sqrt{\cosh ^2(r_{\text{in}})-(1-2 \kappa_1)^2 \sinh ^2(r_{\text{in}})}},\\
d_1^2 &=\frac{\kappa_1^2 e^{-4 r_{\text{in}}} \left(e^{2 r_{\text{in}}}-1\right)^2 \zeta_A^2 \left(e^{2 r_{\text{in}}} \sqrt{\kappa_1 \left(e^{-2 r_{\text{in}}}-1\right)+1}+\sqrt{\kappa_1 \left(e^{2 r_{\text{in}}}-1\right)+1}\right)^2}{\sqrt{\cosh ^2(r_{\text{in}})-(1-2 \kappa_1)^2 \sinh ^2(r_{\text{in}})}},\\
d_2^2 &=\frac{\kappa_1^2 e^{-4 r_{\text{in}}} \left(e^{2 r_{\text{in}}}-1\right)^2 \zeta_A^2 \left(\sqrt{\kappa_1 \left(e^{2 r_{\text{in}}}-1\right)+1}-e^{2 r_{\text{in}}} \sqrt{\kappa_1 \left(e^{-2 r_{\text{in}}}-1\right)+1}\right)^2}{\sqrt{\cosh ^2(r_{\text{in}})-(1-2 \kappa_1)^2 \sinh ^2(r_{\text{in}})}}.
\end{aligned}
}
\end{equation}
Substituting the non-zero coefficients in Eq.~(\ref{Eq:NonZeroCoeffs_QE}) into Eq.~(\ref{Eq:CovQFI_Enhanced}), we find the analytical expression for the covariance-matrix contribution to the QFI to be
\begin{equation}
F_\sigma=4 \kappa_1 \left(\xi_A^2+\zeta_A^2\right) \sinh ^2(r_{\text{in}}).
\end{equation}

Therefore, the QFI for the quantum-enhanced scheme is given by the following expression
\begin{equation}\label{Eq:GaussianQFI_TwoMode}
\mathcal{F}_{d}^{\text{QE}}=F_{\bar{\mathbf{x}}} + F_\sigma = \frac{4 \kappa_0 \eta_A^2 N_{0,\text{q}}}{\kappa_1 e^{-2 r_{\text{in}}}+1-\kappa_1} + 4 \kappa_1 \left(\xi_A^2+\zeta_A^2\right) \sinh ^2(r_{\text{in}}).
\end{equation}
\textit{This equation represents the main result of this work.}
We note that it reduces to Eq.~(51) of~\cite{GaussianQFI2024} in the absence of receiver-aperture truncation and its associated loss.

\subsection{Quantum Advantage under Same-Total-Energy Constraint}\label{Sec:FairQA}
As defined in Sec.~\ref{Sec:QA}, the quantum advantage is given by
\begin{equation}
    Q=\frac{\Delta d_{\text{SQL}}}{\Delta d_{\text{QE}}}=\sqrt{\frac{\mathcal{F}_{d}^{{\text{QE}}}}{\mathcal{F}_{d}^{{\text{SQL}}}}},
\end{equation}
where $\mathcal{F}_{d}^{{\text{SQL}}}$ is given by Eq.~(\ref{Eq:GaussianQFI_SingleMode}) and $\mathcal{F}_{d}^{{\text{QE}}}$ is given by Eq.~(\ref{Eq:GaussianQFI_TwoMode}). To determine the quantum advantage, we impose a fair same-total-energy constraint, under which the total mean number of transmitted photons is the same in the optimal classical scheme and the quantum-enhanced scheme. 
Specifically, for a given coherent-state mean photon number $N_{0,\text{q}}$ in the $\text{HG}_{00}$ mode and squeezing parameter $r_{\text{in}}$ in the $\text{HG}_{10}$ mode of the quantum-enhanced scheme, the corresponding optimal classical scheme is assigned an $\text{HG}_{00}$-mode mean photon number
\begin{equation}\label{Eq:N0cFair}
    N_{0,\text{c}} =N_{0,\text{q}}+\sinh^2{(r_{\text{in}})},
\end{equation}
so that the additional mean photon number $\sinh^2{(r_{\text{in}})}$ contributed by the squeezed-vacuum component is included in the energy budget.

\section{Results}\label{Sec:Results}

\begin{figure*}[htbp!]
	\centering
	\subfloat[$N_{0,\text{q}}=1$ ($\omega_0=5\,\text{cm}$, $a=15\,\text{cm}$). \label{fig:N0_1}]{ 
		\includegraphics[width=0.5\linewidth]{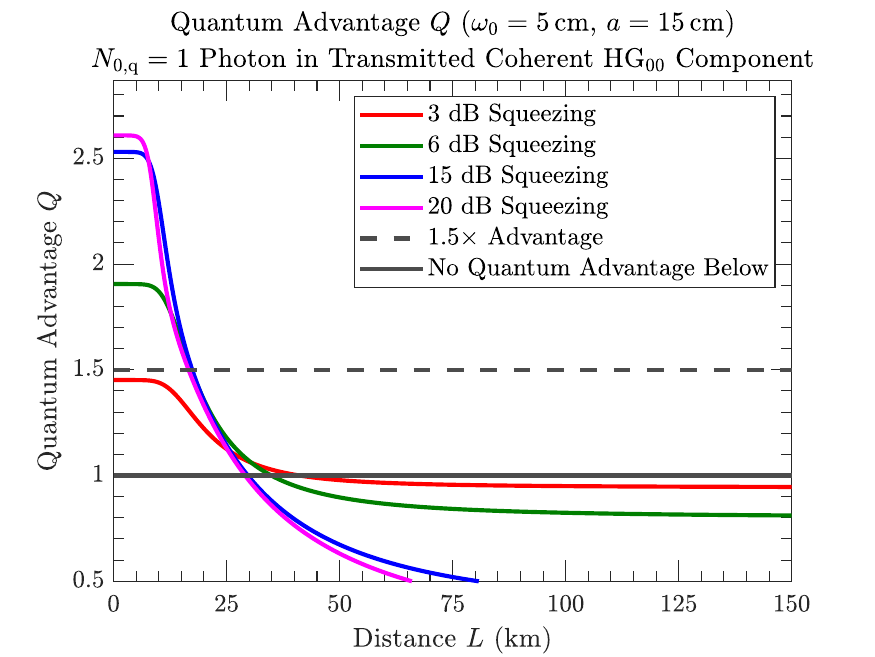}}
	\subfloat[$N_{0,\text{q}}=10$ ($\omega_0=5\,\text{cm}$, $a=15\,\text{cm}$). \label{fig:N0_10}]{ 
		\includegraphics[width=0.5\linewidth]{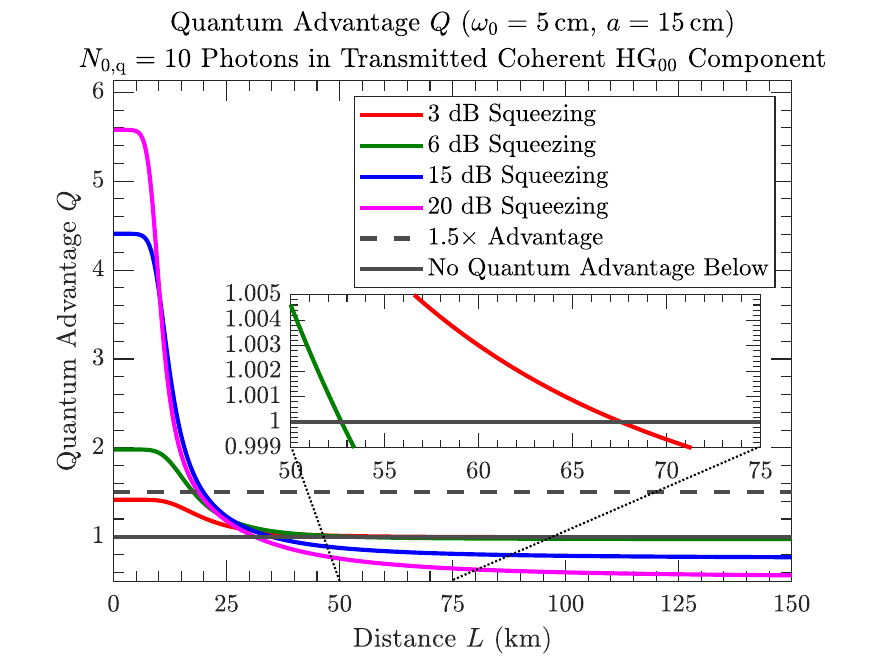}}\\
	\subfloat[$N_{0,\text{q}}=10^2$ ($\omega_0=5\,\text{cm}$, $a=15\,\text{cm}$). \label{fig:N0_100}]{ 
		\includegraphics[width=0.5\linewidth]{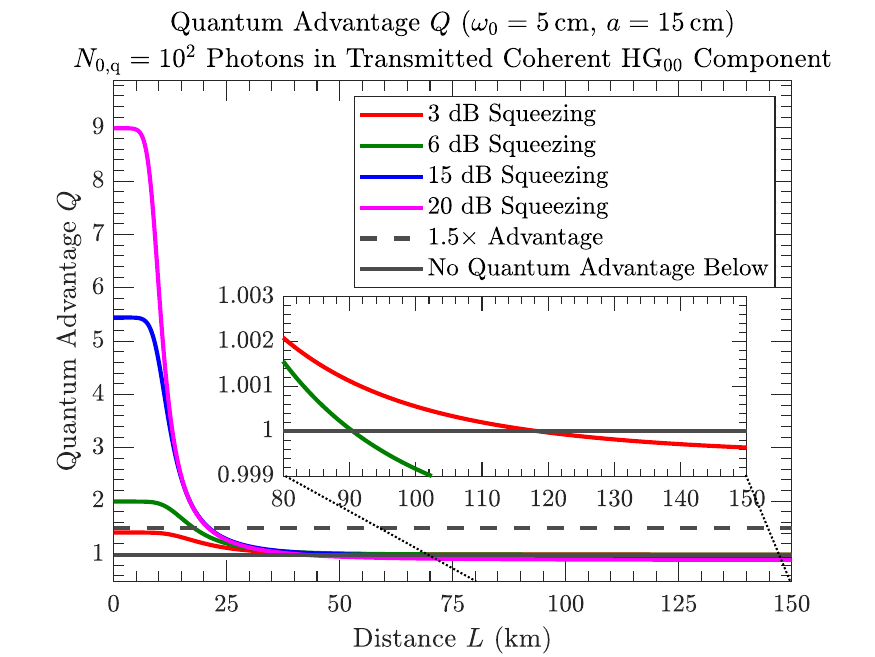}}
	\subfloat[$N_{0,\text{q}}=10^3$ ($\omega_0=5\,\text{cm}$, $a=15\,\text{cm}$). \label{fig:N0_1000}]{ 
		\includegraphics[width=0.5\linewidth]{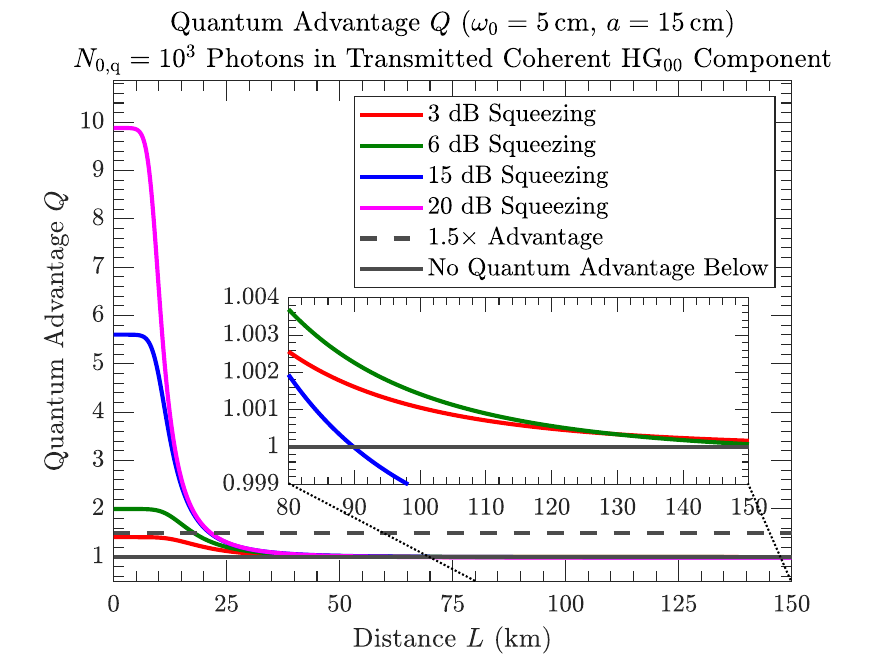}}
	\caption{Quantum advantage $Q$ in displacement estimation for inter-satellite QSTT achieved with different levels of input squeezing expressed as $-10\log_{10} [{\exp(-2r_{\text{in}})}]$ dB. 
    Different subfigures correspond to different mean photon numbers in the coherent HG$_{00}$-mode component of the quantum-enhanced scheme, allowing the quantum advantage to be fairly assessed and compared under different total energy budgets. The results presented in all subfigures are obtained for the typical inter-satellite configuration described in Sec.~\ref{Sec:SysModel}, with $\omega_0=5\,\text{cm}$ and $a=15\,\text{cm}$.}
	\label{fig:Q_vs_N}
\end{figure*}

In Fig.~\ref{fig:Q_vs_N}, we adopt the typical inter-satellite configuration described in Sec.~\ref{Sec:SysModel} (with $\omega_0=5\,\text{cm}$ and $a=15\,\text{cm}$) and plot the quantum advantage \(Q\) in displacement estimation for inter-satellite QSTT as a function of the inter-satellite distance \(L\). Specifically, we consider several representative levels of input squeezing, expressed as \(-10\log_{10}[\exp(-2r_{\text{in}})]\) dB, used in the quantum-enhanced scheme. We note that state-of-the-art squeezing levels can reach approximately 15 dB~\cite{15dB_Squeezing}. Different subfigures correspond to different mean photon numbers \(N_{0,\text{q}}\) in the coherent HG\(_{00}\)-mode component of the quantum-enhanced scheme, allowing the quantum advantage to be fairly assessed and compared under different total energy budgets according to the method described in Sec.~\ref{Sec:FairQA}.

Across all subfigures in Fig.~\ref{fig:Q_vs_N}, we first observe that at the chosen aperture setting a quantum advantage can be achieved over distances of up to $150\,\text{km}$ (primarily at low squeezing levels).
However, the quantum advantage is seen to be a strong decreasing function of the inter-satellite distance.
This behavior arises because squeezing is highly vulnerable to loss, which is caused by diffraction-induced beam spreading followed by truncation by the finite-sized receiver aperture in typical inter-satellite configurations. We also observe that increasing the level of squeezing can reduce the maximum distance over which a quantum advantage is maintained, suggesting that the advantage provided by stronger squeezing is more sensitive to loss. 

When the inter-satellite distance is sufficiently short, higher input squeezing is generally preferred. Indeed, higher squeezing provides a larger quantum advantage in most regimes where a quantum advantage is achievable. 
The main exception occurs in the low-energy regime, where the mean photon number in the coherent HG\(_{00}\)-mode component is small (as shown in Figs.~\ref{fig:N0_1} and~\ref{fig:N0_10}). 
This is because the additional $\sinh^2(r_{\text{in}})$ photons contributed by the squeezed-vacuum HG\(_{10}\)-mode component of the quantum-enhanced scheme become non-negligible under the fair same-total-energy constraint. Since the squeezed-vacuum component is highly vulnerable to loss, it can be more energy-efficient in the low-energy regime to allocate a larger fraction of the photon budget to the coherent HG\(_{00}\)-mode component rather than to the squeezed-vacuum HG\(_{10}\)-mode component. 

In the higher-energy regime, where the mean photon number in the coherent HG\(_{00}\)-mode component is relatively large, the distance-dependent behavior of the quantum advantage becomes almost independent of this number (as shown in Figs.~\ref{fig:N0_100} and~\ref{fig:N0_1000}). The reason for this observation is that, in this high-energy regime, \(N_{0,\text{q}}\gg\sinh^2(r_{\text{in}})\) in Eq.~(\ref{Eq:N0cFair}), so that \(N_{0,\text{c}}\simeq N_{0,\text{q}}\), and the photons associated with the squeezed-vacuum HG\(_{10}\)-mode component become negligible in the total energy budget. In this regime, the quantum advantage is mainly determined by the input squeezing and the inter-satellite distance. In the limit of large inter-satellite distance, the squeezed-vacuum component is degraded by loss towards vacuum, and the quantum-enhanced scheme therefore approaches the SQL-limited performance of the optimal classical scheme. This explains why \(Q\) approaches unity at long inter-satellite distances. This behavior is particularly evident in the red ($3\,\text{dB}$ squeezing) and green ($6\,\text{dB}$ squeezing) curves in Fig.~\ref{fig:N0_1000} and becomes even clearer as $N_{0,\text{q}}$ increases further (results not shown).

Overall, Fig.~\ref{fig:Q_vs_N} shows that although the benefit provided by squeezing rapidly diminishes as the inter-satellite distance increases, a quantum advantage can be achieved over distances of up to $150\,\text{km}$, primarily at low squeezing levels, although $Q\ge 2$ is found only at close range.
We note at Starlink's current deployment of order 10,000 satellites, a physical separation of $100$--$150\,\text{km}$ is maintained for satellites in the same orbital path~\cite{rm10}. As more satellites are deployed by Starlink, and other operators, this inter-satellite distance will reduce.

\begin{figure*}[htbp!]
	\centering
	\subfloat[$N_{0,\text{q}}=10^2$ ($\omega_0=5\,\text{cm}$, $a=30\,\text{cm}$). \label{fig:N0_100_ra30}]{ 
		\includegraphics[width=0.5\linewidth]{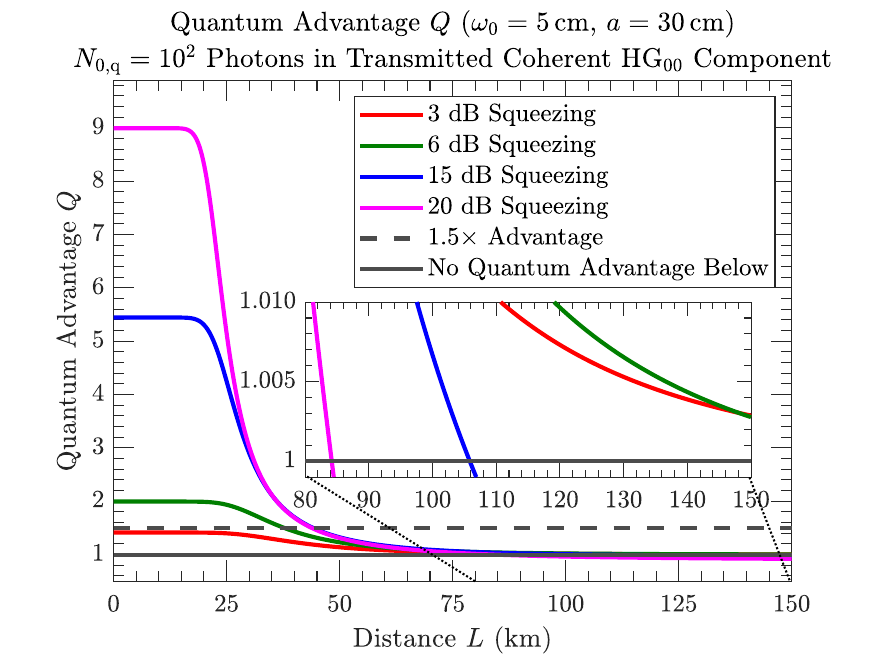}}
	\subfloat[$N_{0,\text{q}}=10^3$ ($\omega_0=5\,\text{cm}$, $a=30\,\text{cm}$). \label{fig:N0_1000_ra30}]{ 
		\includegraphics[width=0.5\linewidth]{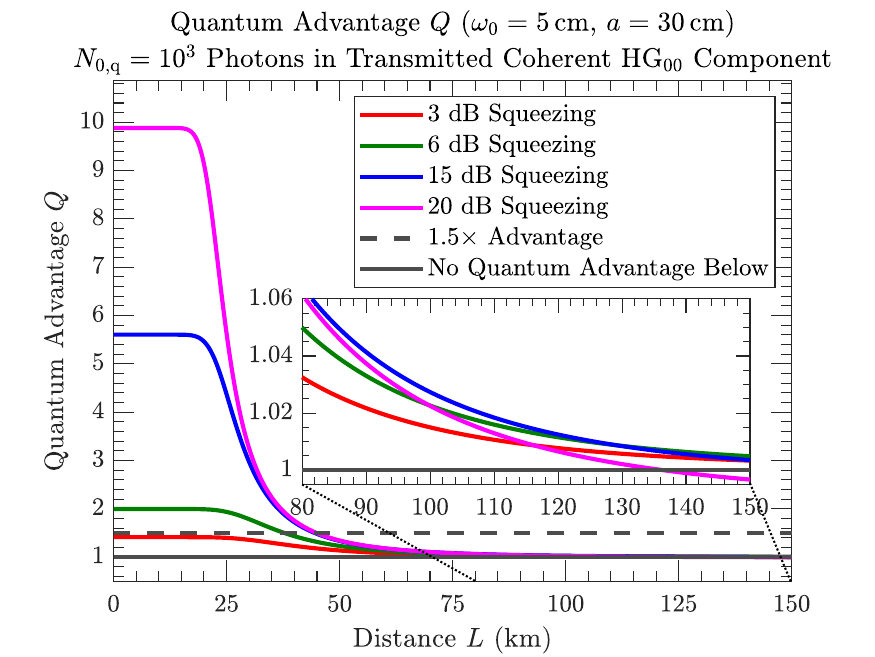}}\\
    \subfloat[$N_{0,\text{q}}=10^2$ ($\omega_0=5\,\text{cm}$, $a=60\,\text{cm}$). \label{fig:N0_100_ra60}]{ 
	   \includegraphics[width=0.5\linewidth]{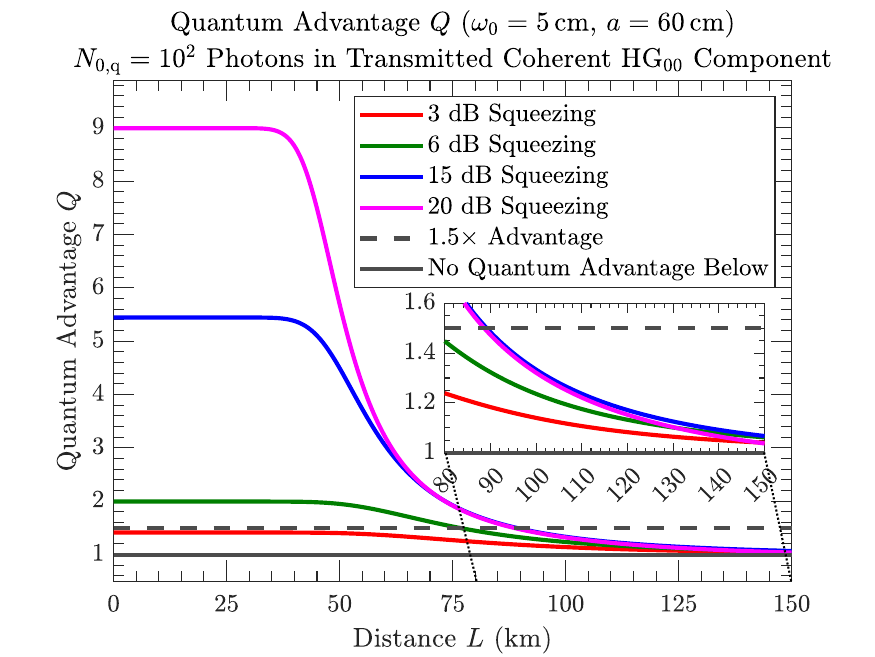}}
	\subfloat[$N_{0,\text{q}}=10^3$ ($\omega_0=5\,\text{cm}$, $a=60\,\text{cm}$). \label{fig:N0_1000_ra60}]{ 
		\includegraphics[width=0.5\linewidth]{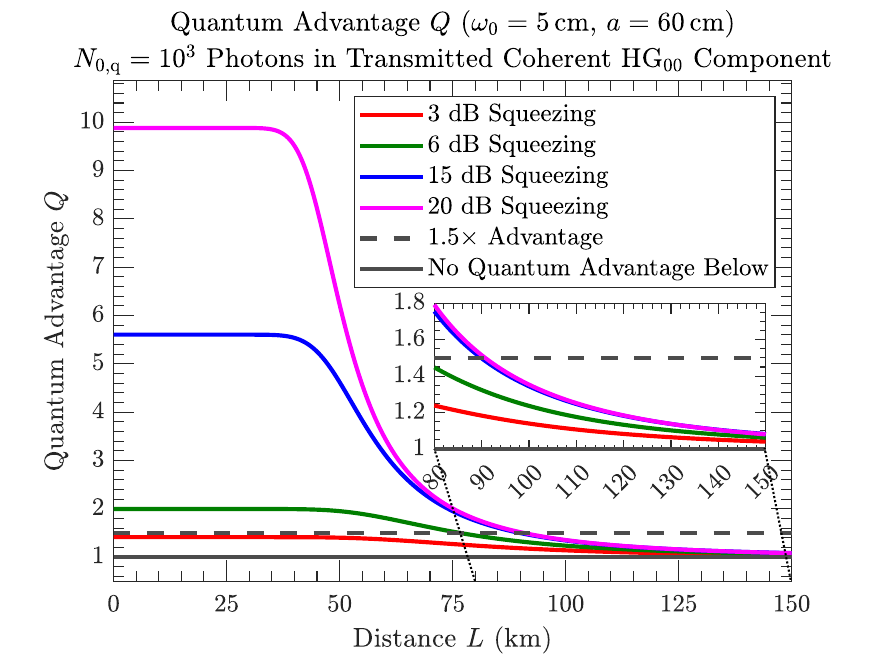}}
	\caption{Quantum advantage $Q$ in displacement estimation for inter-satellite QSTT achieved with different levels of input squeezing expressed as $-10\log_{10} [{\exp(-2r_{\text{in}})}]$ dB. 
    Different subfigures correspond to different mean photon numbers in the coherent HG$_{00}$-mode component of the quantum-enhanced scheme, allowing the quantum advantage to be fairly assessed and compared under different total energy budgets. The results in all subfigures are obtained using the same settings as in Fig.~\ref{fig:Q_vs_N}, except that larger values of the receiver-aperture radius $a$ are used.}
	\label{fig:Q_vs_N_LargeRx}
\end{figure*}

The results in Fig.~\ref{fig:Q_vs_N} indicate that the quantum advantage provided by the use of squeezing is highly sensitive to loss. Reducing loss may therefore allow this advantage to persist over longer distances and become more substantial at a given distance. Advances in optical design and implementation may enable future space platforms to accommodate larger onboard optics. Motivated by this possibility, in Fig.~\ref{fig:Q_vs_N_LargeRx} we examine whether increasing the receiver-aperture radius $a$ can enhance and maintain the quantum advantage over longer inter-satellite distances. 
Specifically, we consider two practical $N_{0,\text{q}}$ values ($10^2$ and $10^3$ photons) relevant to the reference signal in practice and use the same settings as in Fig.~\ref{fig:Q_vs_N}, except that the receiver-aperture radius is increased to $a=30\,\text{cm}$ and $a=60\,\text{cm}$, corresponding to twice and four times the baseline receiver-aperture radius ($a=15\,\text{cm}$), respectively. We again plot the quantum advantage $Q$ in displacement estimation for inter-satellite QSTT as a function of the inter-satellite distance $L$. Comparing the corresponding subfigures in Figs.~\ref{fig:Q_vs_N_LargeRx} and \ref{fig:Q_vs_N}, we observe that increasing the receiver-aperture radius indeed both enhances the achievable quantum advantage at long distances and extends the maximum distance over which it is maintained. A larger receiving aperture also allows stronger squeezing to provide a greater quantum advantage at these distances. In particular, doubling the receiver-aperture radius enables an appreciable quantum advantage over currently practical inter-satellite distances of $100$--$150\,\text{km}$. Quadrupling the radius yields a quantum advantage of approximately $Q=1.5$ at an inter-satellite distance of approximately $90\,\text{km}$.

We close by considering the absolute  scales of the beam displacement present in our results. In the QSTT context, which is the main interest of this paper, we would anticipate the use of very low flux photon beams. Low fluxes are generally the case for single (or entangled) photon sources as used in Discrete-Variable (DV) quantum communications. However, this DV setting would be more appropriate to the use of single photon detectors and the use of  mode sorters. In the case of Continuous-Variable (CV) quantum communications, which this study is more relevant to, the use of a local local oscillator, constructed from very weak reference pulses, rather than a transmitted local oscillator (for enhanced security reasons) also requires low transmitted photon fluxes. At optical wavelengths and for $\sim 100$ photons received, the SQL in our calculations for the beam-displacement estimation is on the order of a few cm, an extremely high accuracy. This uncertainty in position scales into a synchronization error of the order of tens of picoseconds, which is right in the range of interest for advanced synchronization applications. As such, any quantum advantage would be useful. Although not our primary concern here, we note that in the more difficult ground-satellite links, much higher initial photon fluxes (in the reference pulses) would be required due to the much higher losses anticipated (50 dB--70 dB) in order for the SQL to provide similar beam-displacement estimation performance. A trade-off with beam-displacement estimation accuracy and quantum security would then be in place. Details on security issues with local oscillators and reference pulses in the context of CV-QKD can be found elsewhere~\cite{rm11}.

\section{Conclusions}
In this work, we investigated the practical achievability of quantum advantage from the use of non-classical squeezing in beam-displacement estimation for inter-satellite QSTT. We considered a practical squeezing-based quantum-enhanced scheme (in which squeezing is added to the derivative mode of a Gaussian probe beam) and applied the QFI formalism to evaluate the achievable quantum advantage offered by this scheme over the SQL performance of an optimal classical beam-displacement estimation scheme under a fair same-total-energy constraint.
We found that although the benefit provided by squeezing rapidly diminishes as the inter-satellite distance increases due to diffraction-induced beam spreading and subsequent truncation by the finite receiver aperture, a quantum advantage can still be expected over inter-satellite distances below approximately 100~km under typical configurations.

\acknowledgements
The authors thank Dr Ben Sparkes and Dr Oliver Thearle for their comments on this work. The Commonwealth of Australia (represented by the Department of Defence) supported this research through a Defence Science Partnerships agreement.

\appendix

\section{Preliminaries}\label{Sec:Preliminaries}
For completeness we introduce the necessary preliminaries on quantum estimation theory and mode-encoded parameter estimation. Note that the contents presented in this appendix are largely adapted from~\cite{GaussianQFI2024}.

\subsection{Quantum Estimation Theory}\label{Sec:QET}
Quantum estimation theory addresses the general problem of estimating a parameter $d$ encoded in a quantum state $\hat{\rho}_d$ from $M$ independent measurements of a given Positive-Operator-Valued Measurement (POVM) defined by the operators $\{\hat{E}_{\mu}\}$. Given an unbiased estimator $\tilde{d}$ of the parameter $d$, its standard deviation $\Delta \tilde{d}$ is bounded according to the Cram\'er-Rao inequality $\Delta \tilde{d} \ge \frac{1}{\sqrt{M \mathcal{F}_{d, \{\hat{E}_\mu\}}}}$ with the Fisher information $\mathcal{F}_{d,\{\hat{E}_\mu\}}$ given by 
\begin{equation}
    \mathcal{F}_{d, \{\hat{E}_\mu\}}=\sum_{\mu} p(\mu | d)\left[\partial_d \log (p(\mu | d))\right]^2,
\end{equation}
where $p(\mu | d)=\operatorname{Tr}\left[\hat{E}_\mu \hat{\rho}_d\right]$ is the conditional probability of obtaining the result $\mu$ for a given value of $d$. The QFI $\mathcal{F}_{d}$ is defined as the Fisher information optimized over all possible POVMs, that is,
\begin{equation}
    \mathcal{F}_{d}=\max_{\{\hat{E}_\mu\}} \mathcal{F}_{d, \{\hat{E}_\mu\}}.
\end{equation}
The QFI can also be evaluated as
\begin{equation}
\mathcal{F}_{d}=\operatorname{Tr}\left[\hat{\mathcal{L}}_{d}^2 \hat{\rho}_{d}\right],
\end{equation}
where $\hat{\mathcal{L}}_{d}$ is the Symmetric Logarithmic Derivative (SLD) implicitly defined by $2 \partial_{d} \hat{\rho}_{d}=\hat{\mathcal{L}}_{d} \hat{\rho}_{d}+\hat{\rho}_{d} \hat{\mathcal{L}}_{d}$. 
According to~\cite{PhaseSpaceQFI2013,GaussianQPE2015}, when $\hat{\rho}_d$ is an $N$-mode Gaussian state defined by displacement vector $\bar{\mathbf{x}}$ and covariance matrix $\sigma$, the SLD can be expressed as the following function of the quadrature operators
\begin{equation}
\hat{\mathcal{L}}_{d}=L_{d}^{(0)}+\mathbf{L}_{d}^{(1) \top} \hat{\mathbf{x}}+\frac{1}{2} \hat{\mathbf{x}}^{\top} L_{d}^{(2)} \hat{\mathbf{x}},
\end{equation}
where 
\begin{subequations}\label{Eq:LTermsGroup}
\begin{align}
& L_{d}^{(0)}=-\frac{1}{2} \operatorname{Tr}\left[\sigma L_{d}^{(2)}\right]-\mathbf{L}_{d}^{(1) \top} \bar{\mathbf{x}}-\frac{1}{2} \bar{\mathbf{x}}^{\top} L_{d}^{(2)} \bar{\mathbf{x}}, \\
& \mathbf{L}_{d}^{(1)}=\sigma^{-1}\left(\partial_{d} \bar{\mathbf{x}}\right)-L_{d}^{(2)} \bar{\mathbf{x}}, \\
& L_{d}^{(2)}=\frac{1}{2} \sum_{l=0}^3 \sum_{j,k=1}^N \frac{a_{j k}^{(l)}}{\nu_j \nu_k-(-1)^l}\left(S^{\top}\right)^{-1} A_{j k}^{(l)} S^{-1}.
\end{align}
\end{subequations}
In Eq.~(\ref{Eq:LTermsGroup}), $S$ and $\{\nu_j\}$ are the symplectic matrix and the corresponding symplectic eigenvalues resulting from the Williamson decomposition of the covariance matrix $\sigma$, and $a_{j k}^{(l)}=\operatorname{Tr}\left[A_{j k}^{(l)} S^{-1}\left(\partial_{d} \sigma\right)\left(S^{\top}\right)^{-1}\right]$ with $A_{j k}^{(l)}$ being a set of $2N \times 2N$ matrices that are zero everywhere except in the $jk$ block where they are given by $i \sigma_y / \sqrt{2}$, $\sigma_z / \sqrt{2}$, $\mathbbm{1}_2 / \sqrt{2}$, and $\sigma_x / \sqrt{2}$ for $l=0,1,2,3$, respectively.
The QFI for a Gaussian state can be explicitly given by the sum of its displacement-vector contribution $F_{\bar{\mathbf{x}}}$ and covariance-matrix contribution $F_{\sigma}$,
\begin{equation}\label{Eq:GaussianQFI_General}
\mathcal{F}_{d}=F_{\bar{\mathbf{x}}} + F_\sigma,
\end{equation}
where 
\begin{subequations}
\begin{align}
& F_\sigma=\operatorname{Tr}\left[L_{d}^{(2)}\left(\partial_{d} \sigma\right)\right]=\frac{1}{2} \sum_{l=0}^3 \sum_{j,k=1}^N \frac{\left(a_{j k}^{(l)}\right)^2}{\nu_j \nu_k-(-1)^l}, \\
& F_{\bar{\mathbf{x}}}=\left(\partial_{d} \bar{\mathbf{x}}\right)^{\top} \sigma^{-1}\left(\partial_{d} \bar{\mathbf{x}}\right) .
\end{align}
\end{subequations}

The QFI establishes the ultimate precision. When the quantum Cram\'er-Rao bound is saturated, the minimum achievable uncertainty is given by
\begin{equation}
    \Delta d = \frac{1}{\sqrt{M \mathcal{F}_{d}}}.
\end{equation}

\subsection{Mode-Encoded Parameter Estimation}\label{Sec:ModeEncodedPE}
In the problem of displacement estimation, the parameter to be estimated is the displacement $d$, which is encoded in a quantum state $\hat{\rho}_d$ expressed in an $n$-mode spatial basis $\{u_k[d](x,y,z)\}$. Here, $n$ denotes the smallest number of spatial modes necessary to describe the considered state, and the corresponding spatial-mode space is $\mathcal{H}_n=\operatorname{span}\left(\left\{u_k[d](x,y,z)\right\}\right)$. 
Since the spatial-mode basis $\{u_k[d](x,y,z)\}$ may itself depend on the parameter $d$, the $d$-dependence of $\hat{\rho}_d$ arises not only through its moments, but also through the spatial-mode functions $\{u_k[d](x,y,z)\}$.

When $\hat{\rho}_d$ is a Gaussian state, the standard quantum estimation theory introduced in Appendix~\ref{Sec:QET}  can be used to directly address the dependence of $\hat{\rho}_d$ on $d$ through the first-order moment and covariance matrix; however, the $d$-dependence arising from the spatial-mode functions $\{u_k[d](x,y,z)\}$ is not explicitly captured by this treatment.
In this work, we adopt the method of Ref.~\cite{GaussianQFI2024} to further account for this dependence. 
To make the $d$-dependence arising from the spatial-mode functions $\{u_k[d](x,y,z)\}$ explicit in the displacement vector and covariance matrix of $\hat{\rho}_d$, the method constructs a unitary transformation from the parameter-dependent spatial-mode basis $\{u_k[d](x,y,z)\}$ to a parameter-independent spatial-mode basis $\{v_k(x,y,z)\}$. The corresponding transformation rule induces a Gaussian channel that maps Gaussian states to Gaussian states. 
In this way, Ref.~\cite{GaussianQFI2024} provides a complete description of the Gaussian state $\hat{\rho}_d$, in which the $d$-dependence is fully expressed through the displacement vector $\bar{\mathbf{x}}_{\text{I}}$ and covariance matrix $\sigma_{\text{I}}$ under a parameter-independent basis. 
According to~\cite{GaussianQFI2024}, the expression for the covariance-matrix contribution to the QFI is given by
\begin{equation}\label{Eq:CovQFI_General}
\begin{aligned}
    F_\sigma=\frac{1}{2} \sum_{l=0}^3 \sum_{j,k=1}^n & \frac{\left(a_{j k}^{(l)}\right)^2}{\nu_j \nu_k-(-1)^l}\\
   & +\frac{1}{2} \sum_{l=0}^3 \sum_{j=1}^n \sum_{k=1}^m \frac{\left(a_{j, k+n}^{(l)}\right)^2+\left(a_{k+n, j}^{(l)}\right)^2}{\nu_j-(-1)^l},
\end{aligned}
\end{equation}
where $n$ denotes the number of initially populated modes $\{u_k[d](x,y,z)\}$, $m$ denotes the number of orthogonal spatial modes spanning the derivative mode space $\mathcal{H}_{\partial}=\operatorname{span}\left(\left\{\partial_{d} u_k[d](x,y,z)\right\}\right) \backslash \mathcal{H}_n$ (following the notation of~\cite{GaussianQFI2024}), and $\{\nu_j\}$ denotes the symplectic eigenvalues of the covariance matrix $V_n$ of the $n$ initially populated modes in the parameter-dependent basis $\{u_k[d](x,y,z)\}$. 
In Eq.~(\ref{Eq:CovQFI_General}), the coefficients $a_{j k}^{(l)}$ are given by
\begin{equation}
    a_{j k}^{(l)}= \begin{cases}\operatorname{Tr}\left[A_{j k}^{(l)} B_n\right] & 1 \leqslant j, k \leqslant n \\ \operatorname{Tr}\left[\tilde{A}_{j k}^{(l) \top} B_\partial\right] & 1 \leqslant j \leqslant n, n<k \leqslant(n+m) \\ \operatorname{Tr}\left[\tilde{A}_{j k}^{(l)} B_\partial^{\top}\right] & n<j \leqslant(n+m), 1 \leqslant k \leqslant n \\ 0 & j, k>(n+m)\end{cases},
\end{equation}
where $\tilde{A}_{j k}^{(l)}$ are $m \times n$ blocks of matrices $A_{j k}^{(l)}$ and the matrices $B_n$ and $B_{\partial}$ are given by
\begin{equation}\label{Eq:BnMatrix_General}
\begin{aligned}
B_n=S_n^{-1} D_n^{\top}\left(V_n-\mathbbm{1}_{2 n}\right)\left(S_n^{\top}\right)^{-1} &+S_n^{-1}\left(V_n-\mathbbm{1}_{2 n}\right) D_n\left(S_n^{\top}\right)^{-1}\\
&+S_n^{-1}\left(\partial_{d} V_n\right)\left(S_n^{\top}\right)^{-1},
\end{aligned}
\end{equation}
\begin{equation}\label{Eq:BdMatrix_General}
B_{\partial}=D_{\partial}^{\top}\left(V_n-\mathbbm{1}_{2 n}\right)\left(S_n^{\top}\right)^{-1}.
\end{equation}
In Eq.~(\ref{Eq:BnMatrix_General}) and Eq.~(\ref{Eq:BdMatrix_General}), $V_n$ denotes the covariance matrix of the $n$ initially populated modes in the parameter-dependent basis $\{u_k[d](x,y,z)\}$, and $S_n$ denotes the symplectic matrix obtained from the Williamson decomposition $V_n=S_n \nu S_n^T$ with $\nu$ being the diagonal matrix whose diagonal entries are the symplectic eigenvalues of $V_n$.
In Eq.~(\ref{Eq:BnMatrix_General}) and Eq.~(\ref{Eq:BdMatrix_General}), $D_n$ ($D_\partial$) is a $2n\times 2n$ ($2n\times 2m$) matrix constructed using the coefficients $c_{k l}(d)$ ($c^{\prime}_{k l}(d)$) in the expansion
\begin{equation}\label{Eq:CoeffsForDMatrixGeneral}
\begin{aligned}
    \partial_{d} u_k[d](x,y,z)=&\sum_{l=1}^n c_{k l}(d) u_l[d](x,y,z)\\
    &+\sum_{l=1}^m c_{k l}^{\prime}(d) u_l^{\prime}[d](x,y,z),
\end{aligned}
\end{equation}
where $u_l^{\prime}[d](x,y,z)$ can be constructed using arbitrary orthonormalization methods (more details can be found in Appendix A of~\cite{GaussianQFI2024}). In this work, we employ the commonly used Gram-Schmidt procedure for such an orthonormalization (see details in Sec.~\ref{Appendix:GS}).

According to~\cite{GaussianQFI2024}, the displacement-vector contribution to the QFI is given by
\begin{equation}\label{Eq:DispQFI_General}
\Scale[0.9]{
\begin{aligned}
F_{\bar{\mathbf{x}}}= \left(\partial_{d} \bar{\mathbf{x}}_n\right)^{\top} V_n^{-1}\left(\partial_{d} \bar{\mathbf{x}}_n\right)&+\left(\partial_{d} \bar{\mathbf{x}}_n\right)^{\top} V_n^{-1} D_n^{\top} \bar{\mathbf{x}}_n+\bar{\mathbf{x}}_n^{\top} D_n V_n^{-1}\left(\partial_{d} \bar{\mathbf{x}}_n\right) \\
&+\bar{\mathbf{x}}_n^{\top}\left(D_n V_n^{-1} D_n^{\top}+D_{\partial} D_{\partial}^{\top}\right) \bar{\mathbf{x}}_n.
\end{aligned}}
\end{equation}
Substituting Eq.~(\ref{Eq:DispQFI_General}) and Eq.~(\ref{Eq:CovQFI_General}) into Eq.~(\ref{Eq:GaussianQFI_General}), we can evaluate the QFI $\mathcal{F}_d$ for the Gaussian state $\hat{\rho}_d$ (in which displacement $d$ is encoded) and determine the quantum advantage defined in Sec.~\ref{Sec:QA}. It can be seen that the resulting QFI is determined by two sets of ingredients: the displacement vector $\bar{\mathbf{x}}_n$ and covariance matrix $V_n$ of the Gaussian state $\hat{\rho}_d$ in the $n$ initially populated modes $\{u_k[d](x,y,z)\}$, and the spatial-mode functions themselves through the coefficients $c_{k l}[d]$ and $c_{k l}^{\prime}[d]$ in Eq.~(\ref{Eq:CoeffsForDMatrixGeneral}).

\bibliography{references}

\end{document}